\documentclass[usenatbib]{mnras}

\usepackage{natbib}
\usepackage{delarray}
\usepackage{color}
\usepackage{amsmath}
\usepackage{amssymb}
\usepackage{here}
\usepackage{graphicx}
\usepackage[utf8]{inputenc}
\usepackage{float}
\usepackage{epsfig}
\usepackage{tensor}
\usepackage{xspace}
\usepackage{multirow}

\usepackage{newtxtext,newtxmath}

\usepackage[T1]{fontenc}
\usepackage{ae,aecompl}

\newcommand{\mach}{\mathcal{M}}

\newcommand{\be}{\begin{equation}} \newcommand{\ee}{\end{equation}}

\newcommand{\solarmass}{\mathrm{M}_{\rm \sun}}
 
\newcommand{\pc}{\mathrm{pc}}

\newcommand{\dderiv}{\mathrm{d}}

\newcommand{\acknowledgments}{\begin{small}\section*{Acknowledgments}\end{small}}
\newcommand\altaffilmark[1]{$^{#1}$}
\newcommand\altaffiltext[1]{$^{#1}$}

\newcommand{\framework}{GH15\xspace}
\newcommand{\frameworkrad}{GKH16\xspace}
\newcommand{\clustering}{H13\xspace}
\newcommand{\Markfeedback}{K11\xspace}
\newcommand{\myquote}[1]{``#1''}

\usepackage[normalem]{ulem}

\title[Stellar Clustering and Multiplicity with Protostellar Feedback]{Protostellar Feedback in Turbulent Fragmentation: Consequences for Stellar Clustering and Multiplicity}

\author[Guszejnov, Hopkins \&\ Krumholz]{
\parbox[t]{\textwidth}{ D\'avid Guszejnov\altaffilmark{1}\thanks{E-mail:guszejnov@caltech.edu}, Philip F. Hopkins\altaffilmark{1} and Mark R. Krumholz\altaffilmark{2}}
\vspace*{6pt} \\
\altaffiltext{1}{TAPIR, MC 350-17, California Institute of Technology, Pasadena, CA 91125, USA} \\
\altaffiltext{2}{Research School of Astronomy and Astrophysics, The Australian National University, Canberra, ACT 2611 Australia} 
}

\date{To be submitted to MNRAS, \today \vspace{-0.6cm}}
\begin{document}
\maketitle
\label{firstpage}


\begin{abstract}

Stars are strongly clustered on both large ($\sim\pc$) and small ($\sim$binary) scales, but there are few analytic or even semi-analytic theories for the correlation function and multiplicity of stars. In this paper we present such a theory, based on our recently-developed semi-analytic framework called MISFIT, which models gravito-turbulent fragmentation, including the suppression of fragmentation by protostellar radiation feedback. We compare the results including feedback to a control model in which it is omitted. We show that both classes of models robustly reproduce the stellar correlation function at $>0.01\,\pc$ scales, which is well approximated by a power-law that follows generally from scale-free physics (turbulence plus gravity) on large scales. On smaller scales protostellar disk fragmentation becomes dominant over common core fragmentation, leading to a steepening of the correlation function. Multiplicity is more sensitive to feedback: we found that a model with the protostellar heating reproduces the observed multiplicity fractions and mass ratio distributions for both Solar and sub-Solar mass stars (in particular the brown dwarf desert), while a model without feedback fails to do so. The model with feedback also produces an at-formation period distribution consistent with the one inferred from observations. However, it is unable to produce short-range binaries below the length scale of protostellar disks. We suggest that such close binaries are produced primarily by disk fragmentation and further decrease their separation through orbital decay.

\end{abstract}

\begin{keywords}
stars: formation -- stars: binaries: general -- galaxies: star clusters: general -- turbulence -- galaxies: star formation -- cosmology: theory
\vspace{-1.0cm}
\end{keywords}

\section{Introduction}\label{sec:intro}

Star formation (SF) is complex problem that involves nonlinear physics (turbulence, chemistry, gravity, radiation, etc.) on a vast dynamic range. To achieve a deeper understanding of this process a number of simplified theoretical models have been proposed that try to pinpoint the physical processes responsible for individual qualitative features. The most common test of these models is a comparison to the initial mass function (IMF), but that is just one aspect of star formation. It has been long proposed that the stellar clustering and multiplicity properties carry the imprints of the physical processes of star formation (\citealt{Kuiper1935}), making them an ideal secondary test for different star formation models.

It is well known that star-forming regions are highly structured, with stellar positions correlated on a wide range of scales \citet{clustering_Lada,Portegies_young_clusters,Bressert_2010,Gouliermis_2015}. The stellar correlation function has been measured in a wide range (about 5 orders of magnitude in radius) and is found to be rising monotonically on smaller scales in all star clusters (\citealt{Simon1997, Nakajima1998, Hartmann2002, Hennekemper2008, Kraus2008}). Despite the overwhelming observational data and statistical analysis \citep{Bate_1998,Cartwright_2004} there has been little effort to formulate a theoretical understanding of why star formation is clustered. A number of authors have measured the clustering of the stars produced in numerical simulations (e.g., \citealt{Klessen00a, Hansen12a}; see the review by \citealt{SF_big_problems} for further references) and found reasonable agreement with observations, but the physical origin of the result was not completely clear. \cite{Hopkins_clustering} (henceforth referred to as \clustering) was the first to provide a quantitative explanation in terms of the statistics of turbulence. Using the excursion set formalism \clustering showed that the correlation function of \myquote{last crossing objects}\footnote{Smallest self-gravitating structures in a fully developed turbulent medium at a fixed time. They are considered to be the analogues of protostellar cores.} is remarkably similar to that of observed cores, which itself is similar to the correlation function of stars (\citealt{Stanke2006}). However, this model has a significant limitation in that it is calculated at a \emph{fixed time}, so the collapse and further fragmentation of cores is not taken into account; it cannot therefore predict the correlation function of stars, nor their multiplicity statistics.

There is similarly an abundance of observational data about the multiplicity properties of stars (e.g. \citealt{Raghavan2010} for Solar-type stars, \citealt{Burgasser2007} for brown dwarfs; see \citealt{Duchene2013} for a more detailed review). It is generally understood that most multiple star systems either form during the star formation phase through common core fragmentation and protostellar disk fragmentation (\citealt{Tohline2002}) or during the cluster dissolution phase \citep{Kouwenhoven_2010,Parker_Meyer_2014}. Most theoretical work is focused on modeling these processes in detailed numerical studies. Hydrodynamical simulations (e.g. \citealt{Bate09a,Bate12a,Offner2010,Krumholz12a}) have shown good agreement with observed multiplicity statistics and found that radiation feedback is essential. However, these simulations necessarily have limited dynamical range and statistics, of key importance for high-mass stars and long range binaries, and pinpointing the key physics in them is quite challenging.

There has also been significant effort to infer \emph{at-formation} multiplicity properties from observations. Both observations (\citealt{Duchene1999, Kraus2008b, Kraus2011}) and simulations have shown that stars are born in complex, multiple systems that are broken up by dynamical effects (e.g., ejection of stars) causing multiplicity to drop (\citealt{Goodwin2007, Kaczmarek2011}) and the period distribution (commonly referred to as the \emph{binary distribution function}) to shift to shorter periods (\citealt{Kroupa1995,Marks2011}). This can be understood as the result of long range binaries being preferentially broken up by ejection events, which also increase the binding between leftover stars (\myquote{hardening}). This means that to reproduce the present day multiplicity and binary distribution functions the at-formation multiplicity should be of order unity for massive stars, and their period distribution should be flat. These findings, however, have recently been called into question. \cite{Parker_2014} showed that the densities of star forming regions are constant or increasing with time, while \cite{Parker_Meyer_2014} found that an initial distribution of stars with unit multiplicity and an excess of wide binaries will not evolve through N-body processes into a distribution consistent with that observed in field stars.

The aim of this paper is to expand upon the work of \clustering by investigating the features imprinted by isothermal fragmentation and protostellar heating through common core fragmentation in the stellar correlation and multiplicities. This is accomplished by utilizing the MISFIT (Minimalistic Star Formation Including Turbulence) semi-analytical framework described by \citet{guszejnov_feedback_necessity} (hereafter referred to as \frameworkrad), which combines the fragmentation formalism of \cite{TurbFramework} and the protostellar heating model of \cite{Krumholz_stellar_mass_origin} (henceforth referred to as \framework and \Markfeedback) to follow the evolution and collapse of a statistical ensemble of giant molecular clouds (GMCs) down to the protostellar size scale. 

The remainder of this paper is organized as follows. First, in Sec. \ref{sec:method} we briefly outline the MISFIT framework that we use. In Sec. \ref{sec:corr_results} we show that the stellar correlation function is insensitive to both initial conditions and underlying physics and that the predicted 2D correlation function agrees well with observations. In Sec. \ref{sec:multip_results} we show that for low mass stars, turbulent fragmentation mediated by radiation feedback can roughly reproduce the observed multiplicities and mass ratio distribution, and provides qualitative agreement with the expected binary distribution function. However, we also show that protostellar disk fragmentation is necessary to explain the short period tail of the distribution. Finally, in Sec. \ref{sec:conclusions} we summarize our findings and conclude.

\section{Model and Methodology}\label{sec:method}

\subsection{Semi-Analytic Framework}

In this study we use an improved version of the MISFIT semi-analytical framework introduced in \framework and \frameworkrad (see Appendix \ref{sec:corrections} for a detailed description of all changes from the previously published version) which allows us to simulate the evolution and fragmentation of GMC sized clouds at a modest computational cost (compared to full radiation-hydro simulations). This not only allows the rapid exploration of different initial conditions and underlying physics but also enables a statistical analysis as we are able to simulate an ensemble of clouds.

This, of course, comes at the cost of some simple approximations. The main assumption of MISFIT is that density fluctuations in collapsing GMCs are created by turbulence and thus obey \myquote{random walk} statistics (see e.g. \citealt{Hopkins_isothermal_turb}). As the cloud collapses it pumps energy into turbulence (so that virial equilibrium is maintained) as motivated by \cite{Brant_turb_pumping} and \cite{Murray_2015_turb_sim}. Unlike most analytical models MISFIT preserves spatial and temporal information and can be easily expanded with additional physics (e.g. equation of state, angular momentum etc.). We show in Appendix \ref{sec:Bate} that, despite these strong assumptions, our results are roughly in agreement with the detailed radiation hydrodynamical simulation of \cite{Bate12a}.

The simulation starts from a GMC with fully developed turbulence and follows its collapse. The density field of the cloud is resolved on a grid with $N^3$ points and is evolved in Fourier space following a Fokker-Planck equation (see \citealt{general_turbulent_fragment} and \framework). For the bulk of this paper we use $N=32$, and in Appendix \ref{sec:numerical_error} we show that this is sufficient to achieve convergence. Every time a new self-gravitating substructure appears (i.e., the cloud fragments) the code is run recursively for each substructure. When the cloud size reaches the pre-defined relative size scale $R_{\rm min}/R_0$ (the \textit{relative termination scale}) the simulation stops. This termination scale represents the length scale where the initial assumptions break down and the self-similar fragmentation cascade is terminated. 

The primary effect that breaks self-similarity and imposes a scale in our calculations is angular momentum, which leads to the formation of a disk once the object has contracted a certain amount. Disk formation is the natural termination scale. In our model the source of angular momentum is random turbulent motion, which in the supersonic limit means that the distribution of the rotational kinetic energy fraction $\beta=E_{\rm rot}/E_{\rm pot}$ is strongly peaked around a few percent \citep{Burkert_Bodenheimer_rotation_2000}, consistent with the observed distribution of protostellar core rotation rates \citep{Goodman1993}. If one translates this into an angular momentum and assumes that the specific angular momentum of fluid elements is conserved during collapse, the characteristic radius of disk formation is $R_{\rm min}\approx\beta R_0$. In this paper we adopt $R_{\rm min}/R_0=0.01$ as our fiducial value for most calculations, and we explore the sensitivity of the results to our choice in Appendix \ref{sec:numerical_error}.

The initial conditions of the parent clouds are defined by their mass $M_{\rm GMC}$, the sonic length $R_{\rm sonic}$ (the scale at which the turbulent velocity dispersion is equal to the sound speed), the sonic mass $M_{\rm sonic}$ (the minimum self-gravitating mass at the sonic scale), and the termination scale $R_{\rm min}$. All other parameters (e.g. temperature, Mach number) can be derived from these. Moreover, the total mass only affects our results by changing the outer scale of the turbulent cascade, a result we demonstrate in Appendix \ref{sec:numerical_error}, so we shall not discuss it further here. For details about initial conditions and the detailed algorithm see \framework in which a detailed step-by-step guide to the model is provided in Appendix A.

The final output of the simulation is a list of protostars and their initial properties (e.g. mass, velocity, position). As we are not accounting for later dynamical processes, our results only apply \emph{at the time of formation}. The leftover unbound material is assumed to escape.

 \subsection{Implementation of Stellar Feedback}

In this paper we investigate the clustering properties of two classes of models: the case of pure isothermal fragmentation and a model with feedback from protostellar heating based on \Markfeedback. Isothermal turbulence is scale-free (\citealt{McKee_ambipolar_scaling, SF_big_problems}), so we expect no absolute scales in any results (although scales from initial conditions may appear), while the heated model imprints a mass scale that is insensitive to initial conditions (hence there is a peak in the IMF, as shown in \frameworkrad. For comparison we also include some runs where in addition to protostellar heating the gas has a \myquote{stiffening} equation of state (EOS). This means that the gas reacts to compression as a sub-isothermal medium at very large scales, isothermally at intermediate scales and transitions to an adiabatic behavior after reaching a threshold volume density where it becomes opaque to its own cooling radiation. We model this effect using a physically motivated EOS based on \citet{Masunaga_EQS_highgamma_ref} and \citet{Glover_EQS_lowgamma_ref}. In this case the effective polytropic index is depends on the local volume density as
\begin{equation}
\label{eq:gamma_phys_rho}
\gamma_{\rm phys}(\rho) =\begin{cases} 
0.8 &\rho < \rho_{\rm crit,1} \\
1.0 &\rho_{\rm crit,1}<\rho < \rho_{\rm crit,2} \\
1.4 &\rho > \rho_{\rm crit,2}\\
		  \end{cases},
\end{equation}
where we set $\rho_{\rm crit,1 }=5000\,\solarmass/\pc^{-3}$ and $\rho_{\rm crit,2 }=5\times 10^8\,\solarmass/\pc^{-3}$ corresponding to $n_{\rm H_2,\rm crit,1 }\approx 10^{5}\,\mathrm{cm}^{-3}$ and $n_{\rm H_2,\rm crit,2}\approx 10^{10}\,\mathrm{cm}^{-3}$. See GKH16 for more details. 

Our treatment of protostellar radiative feedback is a fairly crude approach motivated by \Markfeedback, and supported numerically by \citet{Krumholz2016}. We assume that the center of self-gravitating clouds collapses first, forming a protostellar seed, then the rest of the cloud accretes onto it. The energy of the matter accreted by this seed is radiated within the optically thick core. The temperature of the material depends on the accretion rate onto the protostar (and thus the mass and dynamical time of the gas around it), and on the energy yield per unit mass from accretion, which we denote $\Psi$. The value of $\Psi$ is set by the protostellar mass-radius relation, and \Markfeedback shows that it is determined primarily by the effects of deuterium burning, which regulates the central temperatures of protostars. Because deuterium burning begins when protostars are only a few $\times 10^{-2}\,\solarmass$, and, for low mass protostars continues for $\sim 10$ Myr, it is the dominant factor in setting $\Psi$ during the bulk of a molecular cloud's star-forming history. Comparing with detailed protostellar evolution calculations, \Markfeedback finds that $\Psi\approx 2.5\times 10^{10}$ J kg$^{-1}$ to better than half a dex accuracy for all protostellar masses in the range $0.05-1\,\solarmass$, and to better than a dex accuracy from $0.01-0.05\,\solarmass$. We therefore adopt this value of $\Psi$ throughout the remainder of this paper. If we assume a spherically symmetric system then ,following \Markfeedback, the gas at $R$ distance from an accreting protostar is heated up to a temperature of
\begin{equation}
T_{\rm heat}^4\approx\frac{\Psi\sqrt{G}}{4\pi\sigma_{\rm SB}}M^{3/2}R^{-7/2},
\label{eq:T_heat}
\end{equation}
where $M$ is the mass enclosed in radius $R$, while $G$, $\sigma_{\rm SB}$ are the gravitational and Stefan-Boltzmann constant respectively. Crudely, this scaling reflects energy conservation as $L=4\pi R^2\sigma_{\rm SB}T_{\rm heat}^4$ for the opaque cloud (see \Markfeedback for more details). Combined, internal heating and the physical processes captured by the EOS of the model set the temperature as
\begin{equation}
\label{eq:T_summing}
T^4=T_{\rm EOS}^4+T_{\rm heat}^4.
\end{equation}
Note that in the feedback only case we use an isothermal EOS which means that $T_{\rm EOS}=T_{0}$ where $T_0$ is the initial temperature of the cloud.
 
It is important to note that protostellar feedback is not scale-free. By using Eq. \ref{eq:T_heat} and assuming virial equilibrium we can find the length scale $\lambda_{\rm heat}$ around a protostar below which heating becomes important ($T_{\rm heat}\geq T_0$):
\be
\lambda_{\rm heat}=\left(\frac{\Psi\sqrt{G}}{4\pi\sigma_{SB}}\right)^{1/2}\left(\frac{k_B}{G \mu m_{\rm H}}\right)^{3/4}T_0^{-5/4}\approx 0.02\,\pc\left(\frac{T_0}{10\,\rm{K}}\right)^{5/4},
\label{eq:lambda_heat}
\ee
where $\mu$ is the mean molecular weight measured in units of $m_{\rm H}$ and $k_B$ is the Boltzmann constant. For our numerical calculations we adopt $\mu=2.3$, appropriate for fully molecular H$_2$ with 1 He per 10 H nuclei. We can similarly find the characteristic mass scale
\be 
M_{\rm heat}\approx 0.5\,\solarmass\left(\frac{T_0}{10\,\rm{K}}\right)^{-1/4}
\label{eq:mass_heat}
\ee
that sets the peak of the IMF (see \Markfeedback for a more detailed calculation that leads to $M_{\rm heat}\propto T_0^{-1/18}$).

To easily identify the results for different models and parameters we use the labels shown in Table \ref{tab:simparam}. The \textit{STD} label refers to initial conditions similar to Milky Way GMCs, while \textit{ULIRG} runs have the very high temperature and strong turbulence characteristic to the clouds of Ultra Luminous Infrared Galaxies (ULIRGs). There are also a number of runs where the physical parameters are not varied but the numerical ones are, so that we can identify numerical artifacts in our results.

\begin{table*}
\footnotesize
	\centering
		\begin{tabular}{ | c | c | c| c | c | c| c |c |c |c |c |c |}
		\hline
		\multirow{2}{*}{\bf Label} & \multicolumn{5}{c|}{\bf Input Parameters}  & \multicolumn{3}{c|}{\bf Derived Parameters}  &  \multirow{2}{*}{\bf Thermodynamics}  \\
		\cline{2-9}
		& \bf $\mathbf M_{\rm GMC}$ [$\solarmass$]& \bf $\mathbf R_{\rm min}/R_0$& \bf N & \bf $\mathbf M_{\rm sonic}$ [$\solarmass$]& \bf $\mathbf R_{\rm sonic}$ [$\pc$] & \bf $\mathbf T_0$ [K] & \bf $\mathbf R_0$ [$\pc$] & \bf $\mathbf \mach_0$ &   \\
		\hline
		\hline
		Isohermal - MW & $10^4$ & $10^{-2}$ & 32 & 1.6 & 0.1 & 10 & 11.1 & 10.5 & Isothermal  \\
		\hline
		Isothermal\_SmallR & $10^4$ & $10^{-3}$ & 32 & 1.6 & 0.1 & 10 & 11.1 & 10.5 & Isothermal  \\
		\hline
		Isohermal - ULIRG & $10^4$ & $10^{-2}$ & 32 & 0.31 & 0.0026 & 75 & 0.66 & 13.1 & Isothermal  \\
		\hline
		\hline
		Heating - MW & $10^4$ & $10^{-2}$ & 32 & 1.6 & 0.1 & 10 & 11.1 & 10.5 & Protostellar Heating  \\
		\hline
		Heating - ULIRG & $10^4$ & $10^{-2}$ & 32 & 0.31 & 0.0026 & 75 & 0.66 & 13.1 & Protostellar Heating  \\
		\hline
		Heating\_N16 & $10^4$ & $10^{-2}$ & 16 & 1.6 & 0.1 & 10 & 11.1 & 10.5 & Protostellar Heating  \\
		\hline
		Heating\_N64 & $10^4$ & $10^{-2}$ & 64 & 1.6 & 0.1 & 10 & 11.1 & 10.5 & Protostellar Heating  \\
		\hline
		Heating\_M1E3 & $10^3$ & $10^{-2}$ & 32 & 1.6 & 0.1 & 10 & 3.5 & 5.2 & Protostellar Heating  \\
		\hline
		Heating\_M1E5 & $10^5$ & $10^{-2}$ & 32 & 1.6 & 0.1 & 10 & 35.4 & 16 & Protostellar Heating  \\
		\hline
		Heating\_SmallRmin & $10^4$ & $10^{-3}$ & 32 & 1.6 & 0.1 & 10 & 11.1 & 10.5 & Protostellar Heating  \\
		\hline
		Heating\_LargeRmin & $10^4$ & $10^{-1}$ & 32 & 1.6 & 0.1 & 10 & 11.1 & 10.5 & Protostellar Heating  \\
		\hline
		\hline
		Heating+EOS - MW & $10^4$ & $10^{-2}$ & 32 & 1.6 & 0.1 & 10 & 11.1 & 10.5 & Heating+EOS  \\
		\hline
		Heating+EOS - ULIRG & $10^4$ & $10^{-2}$ & 32 & 0.31 & 0.0026 & 75 & 0.66 & 13.1 & Heating+EOS  \\
		\hline
		\end{tabular}
	\caption{Initial conditions of the different models presented in this paper. The actual input parameters of our models are the GMC mass $M_{\rm GMC}$, sonic mass $M_{\rm sonic}$ and length $R_{\rm sonic}$, from which more physical parameters like initial temperature ($T_0$), radius ($R$) and Mach number ($\mach_0$) at the cloud scale ($R=R_{\rm GMC}$) can be derived. The resolution of each \myquote{refinement level} of our semi-analytic calculation is set by $N^3$: but each structure is continuously followed and sub-refined. This cascade is terminated when a cloud reaches $R_{\rm min}/R_0$ relative size, the cascade \myquote{termination scale}, without fragmentation. All calculations were performed for a statistical ensemble ($20-200$) of GMCs.}
	\label{tab:simparam}
\end{table*}

\subsection{Clustering and Multiplicity Statistics}
\label{ssec:clusterstat}

For each simulation we have as output a list of stellar masses and positions. From these, we compute several statistical quantities describing the stellar distribution. Our first quantity of interest is the correlation function. In this paper we adopt the usual definition of the 3D correlation function $\xi(r)$,
\begin{eqnarray}
P(r,dr)=\frac{\langle N(r,\dderiv r)\rangle}{\langle n \rangle\, \dderiv V}\nonumber\\
1+\xi(r)=\lim\limits_{\dderiv r \to 0} P(r,dr),
\label{eq:correlation}
\end{eqnarray}
where $N(r,\dderiv r)$ is the number of objects whose distance is $\in [r,r+\dderiv r]$, $n$ is the density of objects, $\dderiv V=4\pi r^2\dderiv r$ and $\langle ... \rangle$ denotes ensemble averaging.

We can similarly define the 2D correlation function $\xi_{\rm 2D}(r)$, which is identical to $\xi$ except that one computes the distance only in 2 of the 3 orthogonal directions. Unlike $\xi$, the stellar $\xi_{\rm 2D}(r)$ is measurable and it is easy to show that $\xi_{\rm 2D}(r)\propto \Sigma_{*}(r)$, where $\Sigma_{*}(r)$ is the mean surface density of stars measured in an annulus at $r$ distance from other stars. For the purpose of generating quantities that can be readily compared to observations, we must also account for sensitivity limits, which make it difficult to detect low mass objects. Since studies of stellar correlation have been performed with a wide range of sensitivities, we simply choose a roughly representative limiting mass $M_{\rm min} = 0.5\,\solarmass$, and compute our correlation function using only stars more massive than this limit.

Our second characteristic of interest is the multiplicity properties of the stars -- both the multiplicity fractions and the distribution of periods and mass ratios. Since our calculations involve no dynamical evolution after the protostars are formed, deriving these statistics is not trivial, as the newly formed stars form a fractal-like structure where each star is bound to a number of other stars. Such a configuration is expected for young star clusters based on simulations, and is completely consistent with the observed distribution of newly-formed stars \citep[e.g.,][]{Kruijssen09a, Bate_lowmass_feedback, Krumholz12a}. However, it makes identification of distinct, bound systems difficult, and leads to structures which are very unlikely to survive for even a single cluster crossing time (e.g. non-hierarchical quadruple systems orbiting each other). Thus it is important that we try to correct for this behavior. In this paper we use the hierarchical algorithm introduced by \cite{Bate_lowmass_feedback}, which has the following steps:
\begin{enumerate}
\item Calculate the binding energy between all pairs of stars.
\item Find the most bound pair and replace it with a single point mass with the same total mass and momentum, located at the center of mass of the removed pair.
\item Recursively repeat steps 1 and 2 until no more bound stars are left, with the exception that we do not combine pairs of objects if the resulting bound aggregate would consist of more than 4 individual stars. If such an aggregate is the most bound pair at any point, proceed to the next most bound pair, terminating if no other bound pair exists. Also, in order to get stable, hierarchical multiples we require that the period of a newly assigned star is at least ten times higher than that of the original aggregate.
\end{enumerate}
This algorithm provides a list of single, binary, triple and quadruple star systems with which we can calculate the \emph{multiplicity fraction} $f$, defined as
\be
\label{eq:multiplicity}
f(M)=\frac{B+T+Q}{S+B+T+Q},
\ee
where $S,B,T,Q$ are the number of single, binary, triple, quadruple systems within which the most massive star (\emph{primary star}) has mass $M$. This definition has the advantage that it can be observed fairly robustly (\citealt{Hubber_Whitworth_2005,Bate_lowmass_feedback}), as this does not differentiate between the classes of multiple star systems, so $f$ does not change if a new companion star is discovered in a binary system. Note that to account for the decreased sensitivity of observations to very low mass stars we neglect companions below $0.01\,\solarmass$.

\section{Results}

\subsection{The Stellar Correlation Function}\label{sec:corr_results}

\begin{figure}
\begin {center}
\includegraphics[width=\linewidth]{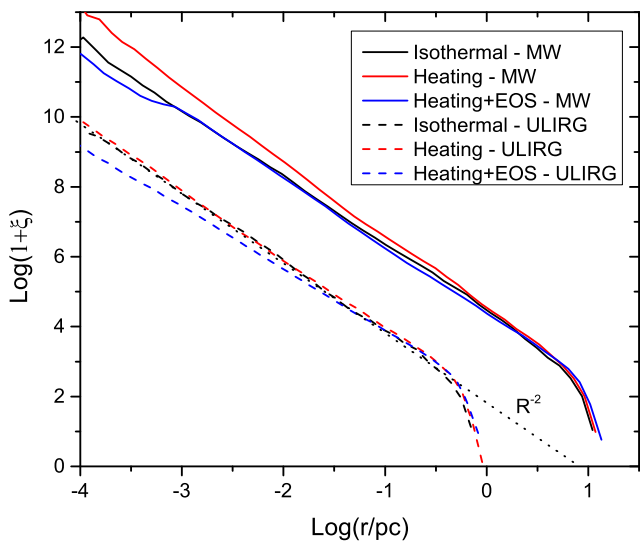}
\caption{
Stellar correlation function for star formation models with isothermal physics (\textit{Isothermal}), protostellar heating (\textit{Heating}) and protostellar heating with an artificial EOS (\textit{Heating+EOS}) for two sets of initial conditions:  the MW-like \textit{MW} and the more extreme \textit{ULIRG}. It is apparent that the initial conditions and underlying physics have limited effect on the qualitative behavior which is close to a power law (the sub-isothermal regime of the EOS models cause a slight difference in the slope). The different large scale cutoffs are introduced by the different initial cloud sizes, and the different normalization simply results from the different linewidth-size relation between the Milky Way and ULIRG cases.}
\label{fig:allstar_corr_comp}
\end {center}
\end{figure}

Figure \ref{fig:allstar_corr_comp} shows the predicted stellar correlation function for a selection of our models, computed using an ensemble average of the $\sim 20-200$ GMCs we have run for each case. The shape is close to a power-law,  $\xi(r)\propto r^{-\gamma}$ with $\gamma\simeq 2$, with a cutoff at the size scale of the parent GMC. These properties are remarkably robust to changes in initial conditions and even to changes in the underlying small scale physics. 

Qualitatively, the isothermal pure-power-law behavior can be understood with a simple toy model: small objects form after significant contraction and a number of fragmentation events for which the physics is self-similar in the isothermal case. So imagine that a cloud contracts by a factor of $\epsilon$, then fragments into two equal-mass fragments. Then each of these two fragments contracts and produces two more fragments, and so on. This prescription is similar to a well-studied fractal, the Cantor Set (Cantor Dust more specifically). For this, the correlation function is a power-law with slope $f(\epsilon)\sim 2$ for $\epsilon\sim 1/2$ (see Appendix \ref{sec:cantor}).

Figure \ref{fig:observations} compares our predictions to the observed surface density of stars (proportional to the 2D correlation function). In examining this plot, note that the absolute values of the stellar surface densities are not meaningful, since these are just dictated by a our choice of sonic length, and thus can be tuned freely by considering slightly different physical scalings, exactly as one might expect when considering a range of star-forming regions of widely varying density, mass, and velocity dispersion. Instead, the meaningful comparison is the shapes of the functions. In this regard, we see that the simulated correlation functions have a slope quite similar to the observations at scales larger than $\sim 10^{-2}\,\pc$. Below this scale our models cannot reproduce the significant steepening of the correlation function. We show below that this directly manifests in the distribution of short period binaries where the simulation fall short of observations at the same scale. This is the length scale of the largest protostellar disks, which suggests that disk physics (which are neglected in these models) is responsible for the steepening. However, we must stress that dynamical relaxation does affect the observed, finite age systems and is probably responsible for outlier systems like Trapezium and Upper Sco. Both of these systems are \textit{dynamically} older, in the sense that they have existed for more crossing times, than the other systems shown, which supports this conjecture. One should be careful not to draw the false conclusion that this model fully explains the observed stellar spatial distribution simply because it reproduces the correlation function. It has been shown that very different geometries (e.g., fractal vs spherical) can lead to similar correlation function slopes \citep{Gouliermis_2014}. Nevertheless we can say that this model is at least consistent with the observed stellar correlation function in the large scale, fractal-like regime.

\begin{figure}
\begin {center}
\includegraphics[width=\linewidth]{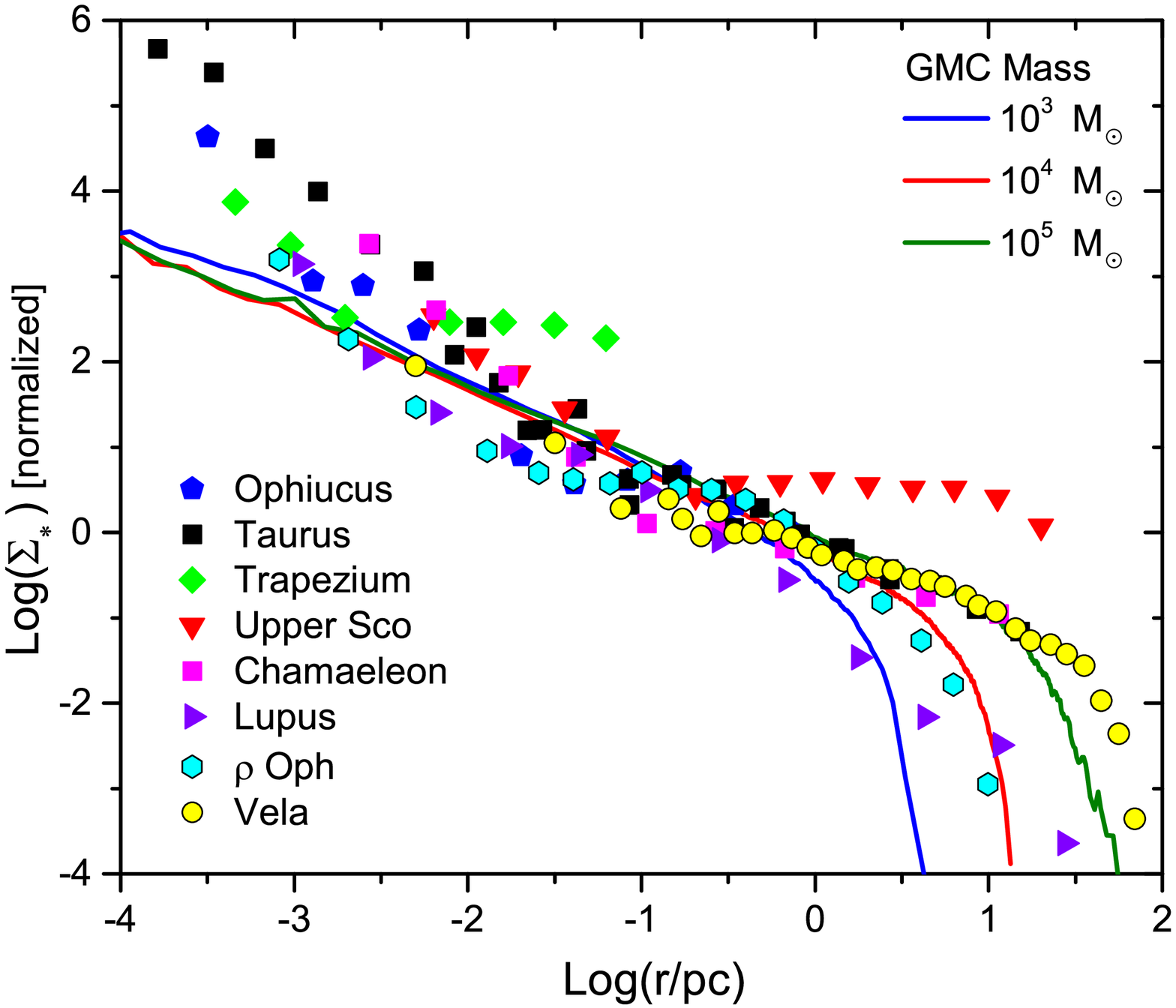}
\caption{Observed surface density of neighboring stars ($\Sigma_*$, which is proportional to the projected correlation function $\xi_{\rm 2D}$) for Chamaeleon, Ophiucus, $\rho$ Oph, Taurus, Trapezium, Upper Sco, Lupus and Vela (using data from \citealt{Simon1997, Nakajima1998, Hartmann2002, Hennekemper2008, Kraus2008}) compared to predicted $\Sigma_{*}$ functions for our models including protostellar heating (solid lines,  \textit{Heating\_M1E3}, \textit{Heating-MW}, \textit{Heating\_M1E5}). The absolute values of the observation depend on a number of external factors so they are normalized to roughly match simulations in the 0.1-1 pc range.}
\label{fig:observations}
\end {center}
\end{figure}

\subsection{Multiplicity} \label{sec:multip_results}

After grouping stars into bound systems following the procedure described in Section \ref{ssec:clusterstat}, we assign each star one of the following labels:
\begin{enumerate}
	\item \textit{Single}: The star is not bound to any other stars.
	\item \textit{Multiple}: The star is the most massive (primary) star of a multiple star system.
	\item \textit{Non-primary}: The star is part of a multiple star system, but it is not the primary star.
\end{enumerate}

Not all of the companion stars that emerge from our analysis would be detectable by current techniques. In particular, brown dwarf companions to main sequence stars are quite hard to detect. Therefore we must correct for completeness before comparing to observations. As a guide to the current observational capabilities, we follow the summary given in Table 8 of \cite{DeRosa_2014_VAST3}. Based on this summary, we apply the following cuts to our data: 
\begin{enumerate}
\item For primaries with mass $M>0.08$ $M_\odot$, we discard any companions with masses below $0.08$ $M_\odot$
\item For primaries with mass $M < 0.08$ $M_\odot$, we discard companions for which the secondary to primary mass ratio is $<0.2$. 
\end{enumerate}
While these cuts are only an approximate representation to the diversity of observational survey selection functions in the literature, they provide a reasonable approximation to the capabilities of the current state of the art.
 
\begin{figure*}
\begin {center}
\includegraphics[width=0.475\linewidth]{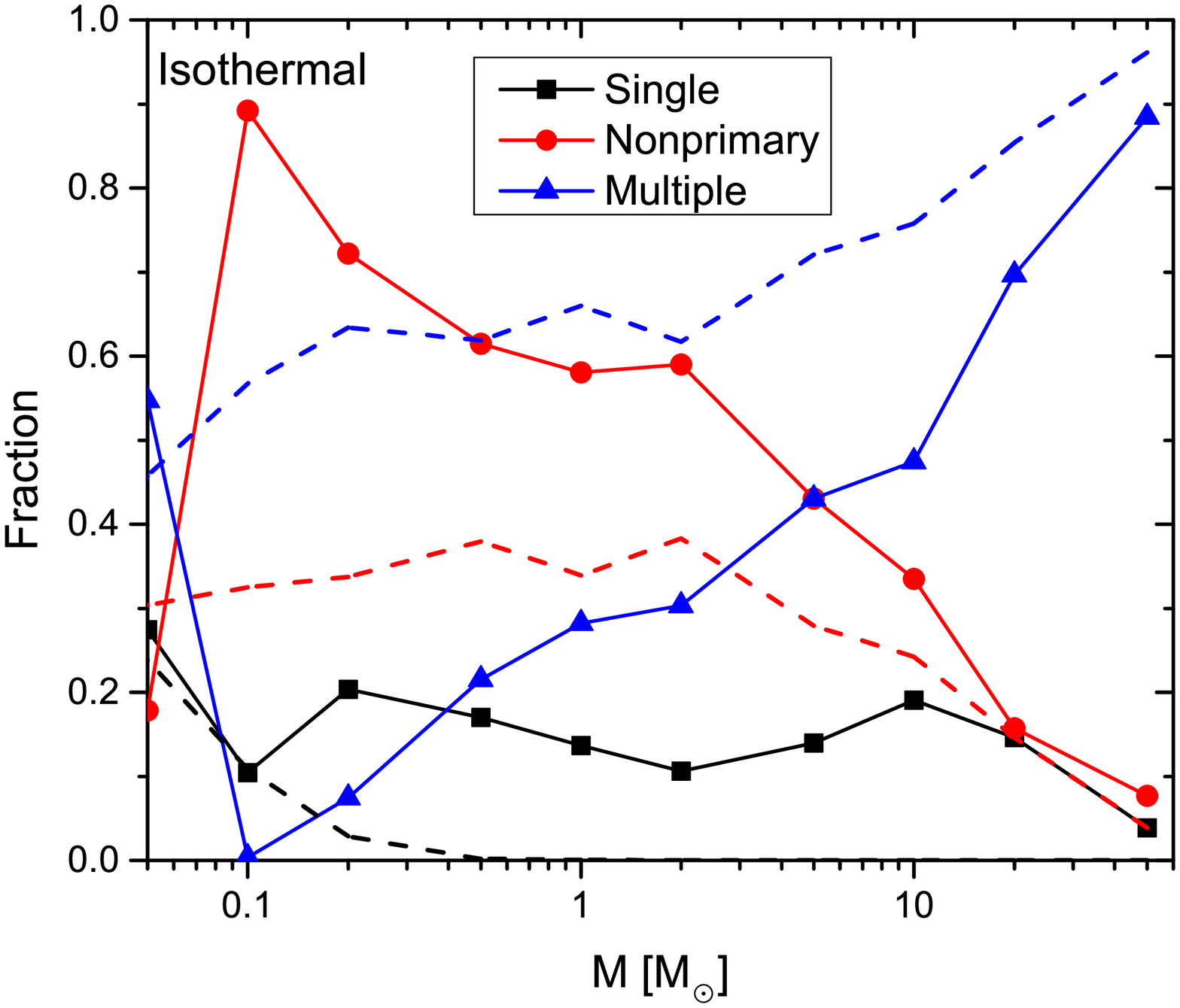}
\includegraphics[width=0.475\linewidth]{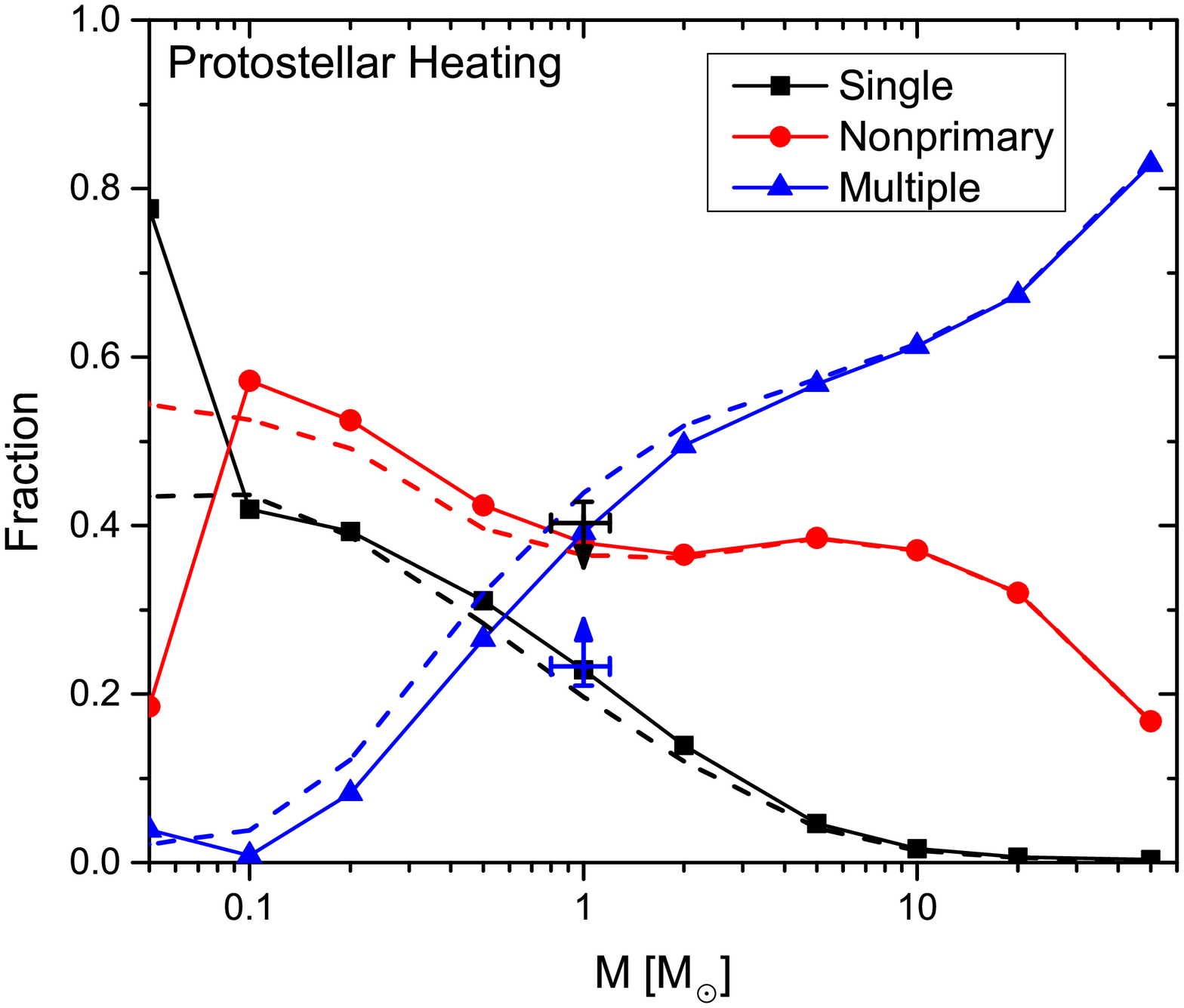}
\caption{Fraction of stars in bound systems as a function of mass in our models of isothermal fragmentation (left; \textit{Isothermal - MW}) and including protostellar heating (right; \textit{Heating - MW}). We assign each star one of the \textit{single}, \textit{multiple} and \textit{nonprimary} labels. The dashed lines show the true distribution predicted by the model, while the solid lines show the results that would be observed given the completeness limits of current surveys. The brackets show the results of \protect\cite{Moe_DiStefano2016} based on their analysis of the \protect\cite{Raghavan2010} observations. The sudden change around $0.1\,\solarmass$ is due to the different observational bias for very low mass stars; see the main text for details.}
\label{fig:Star_status}
\end {center}
\end{figure*}

Fig. \ref{fig:Star_status} shows the fraction of stars in each of the three classes as a function of stellar mass before and after applying the observational bias. Since the isothermal model has no inherent physical scale, in an ideal case we would not expect any mass dependence. However, the finite initial mass $M_{\rm GMC}$ and the cutoff imposed by observational selection affect the results. The former leads to finite size effects at larger masses. Specifically, since there is a finite total mass, there must be a single most massive star, and for obvious reasons it cannot be non-primary. Similarly, other stars that are near the most massive are also biased against being non-primary. This effect is responsible for the decline in the non-primary curve at high masses. At the other end of the mass spectrum, the fact that brown dwarf companions to hydrogen-burning primaries are difficult to detect explains the sharp decline in the non-primary fraction and sharp rise in the multiple fraction for the lowest mass bin. The sharp change in behaviour above and below $0.1\,\solarmass$ has a simple explanation: for hydrogen-burning stars, multiplicity surveys are primarily conducted in the field, while for brown dwarfs they are mainly conducted in young clusters. Since brown dwarfs are easier to detect in young clusters than in the field, surveys of brown dwarf primaries are much more complete in finding brown dwarf companions than surveys of hydrogen-burning primaries.

The case including stellar radiation feedback looks qualitatively similar to the isothermal one. In both cases we recover the simple rule that more massive stars tend to be the primary stars of systems while smaller stars tend to be their companions. However, most of the stars in the isothermal model were born in systems of multiple stars, while there is a significant number of single stars in the radiative heating case. The transition from where it becomes more common for stars to be the primaries of multiple systems than to be single is $\sim 1$ $M_\odot$ which is a result of the peak in the stellar mass function imposed by heating, which suppresses the formation of stars below the IMF peak at $\sim 0.3\,\solarmass$. Smaller stars are unlikely to be primaries mainly because there are increasingly few lower mass stars available to be their companions.

We can also compare the results of our models to observations. In Fig. \ref{fig:Star_status} we plot in the right panel the results of \citet{Moe_DiStefano2016}, based on analysis of the observations of \citet{Raghavan2010}. Compared to these observations, our model slightly overpredicts the multiplicity of solar type stars and underpredicts the fraction that are single. 

\begin{figure}
\begin {center}
\includegraphics[width=\linewidth]{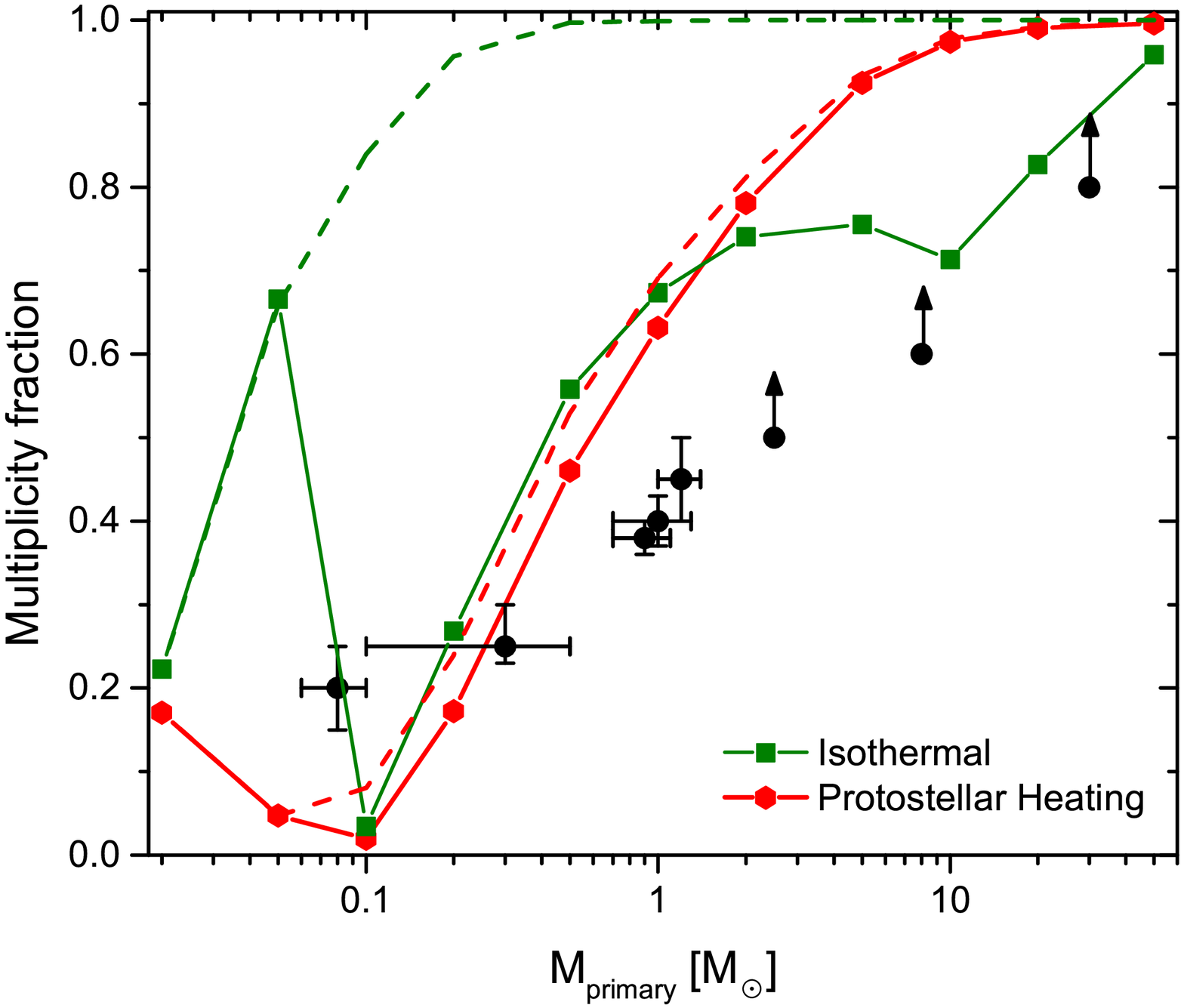}
\caption{
The multiplicity of stars of different masses in the isothermal case (\textit{Isothermal - MW}) and the model with protostellar heating (\textit{Heating - MW}) compared to the observed multiplicity fractions (black circles with error bars) from the review of \protect\cite{Duchene2013}. The dashed lines show the results without the observational bias. Both models overpredict the multiplicity fraction as dynamical processes are neglected in our simulations, but the effect is far more severe for the isothermal model, particularly at low masses. Note that the sudden change around $0.1\,\solarmass$ is due to the different observational bias for very low mass stars.}
\label{fig:multiplicity_fraction}
\end {center}
\end{figure}

We compare the stellar multiplicity as a function of primary mass with observations in  Figure \ref{fig:multiplicity_fraction}. We find that the observed results for the heated and isothermal cases are similar, and both qualitatively reproduce the observational result that the multiplicity fraction is near unity for primaries substantially above $1$ $\solarmass$, dropping to tens of percent for $\sim 0.3\,\solarmass$ or smaller primaries. However, the apparent similarity between the observed distributions for the isothermal and heated cases is primarily an illusion due to observational completeness effects. In the isothermal case, essentially \textit{every} $\sim 1\,\solarmass$ primary has an undetected brown dwarf companion, and thus the true multiplicity fraction for primaries of this mass is close to unity. It is only our inability to detect these brown dwarfs that makes the predicted distribution in the isothermal case at all compatible with the observations.

It is also worth investigating how these results would be affected by protostellar disk fragmentation. To do so, we construct a toy model for protostellar disk fragmentation that can be used to post-process our simulation results, based on the works of \cite{Kratter10a} and \cite{Offner2010}. They define the thermal parameter of the disk as $\xi=\dot{M}_{\rm in}G/c_{s,d}^3$, where $\dot{M}_{\rm in}$ is the infall accretion rate onto the disk and $c_{s,d}$ is the sound speed of the disk. Both find that $\xi$ is the main parameter in determining whether a protostellar disk fragments or not. Physically, $\xi$ is just the ratio of the accretion rate $\dot{M}_{\rm in}$ into the disk to the maximum rate at which a gravitationally stable disk with dimensionless viscosity $\alpha\lesssim 1$ can deliver mass to the central star, which is $\sim c_{s,d}^3/G$. 

Using our protostellar heating prescription we can express $c_{s,d}^2\propto T_d \approx T_{\rm heat} \left(R_d/R_{\rm core}\right)^{-1/2}$ where we have used that $T^4\propto R^{-2}$ in an opaque medium. As noted above, the outer edge of the disk should be found at a radius $R_d \approx \beta R_{\rm core}$, where $\beta$ is the rotational kinetic energy divided by the binding energy. Using Eq. \ref{eq:T_heat} to evaluate $T_d$ at this radius $R_d$ yields
\be
\xi\sim \left(\frac{G^{7/8} \mu m_{\rm H} (4\mathrm{\pi}\sigma_{\rm SB})^{1/4}}{\Psi^{1/4} k_b}\right)^{3/2}\beta^{3/4} R_{\rm core}^{-3/16} M^{15/16},
\ee
where we used $c_s^2 = k_B T / \mu m_{\rm H}$ with $\mu m_{\rm H}$ as the molecular weight of the gas. $\xi$ has a weak dependence on the radius so we can safely use the $R\sim 10^{-4}\,\pc$ protostellar disk size scale. This leads to $\xi\approx 0.65 \left(M/\solarmass\right)^{15/16}$. Based on Fig. 2 of \cite{Kratter10a} fragmentation is very likely if $\xi>1$, which corresponds to collapsing final fragment masses $>1.5\,\solarmass$ in our model. Thus, in this crude approximation, the only effect protostellar disk fragmentation would have on our multiplicity fraction in Fig. \ref{fig:multiplicity_fraction} is that it would reach unity at a somewhat lower stellar mass. Our conclusion that low mass disks are for the most part too warm to fragment, but that disk fragmentation should be common for somewhat super-Solar and larger stars, is consistent with the numerical results of \citet{Offner2010}.

\subsection{Demographics of the Binary Population}

\subsubsection{Mass Ratios and the Brown Dwarf Desert}

\begin{figure*}
\begin {center}
\includegraphics[width=0.475\linewidth]{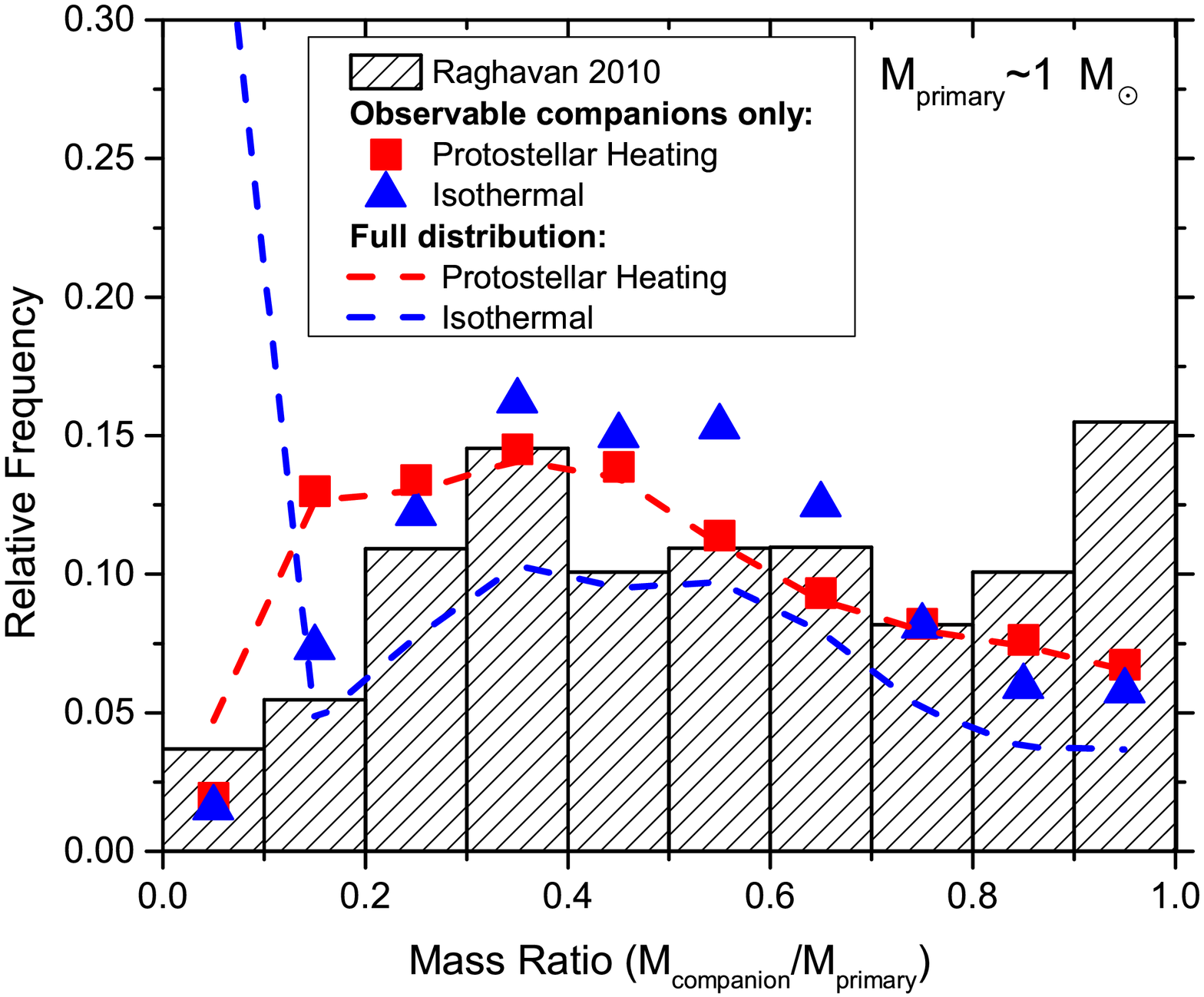}
\includegraphics[width=0.475\linewidth]{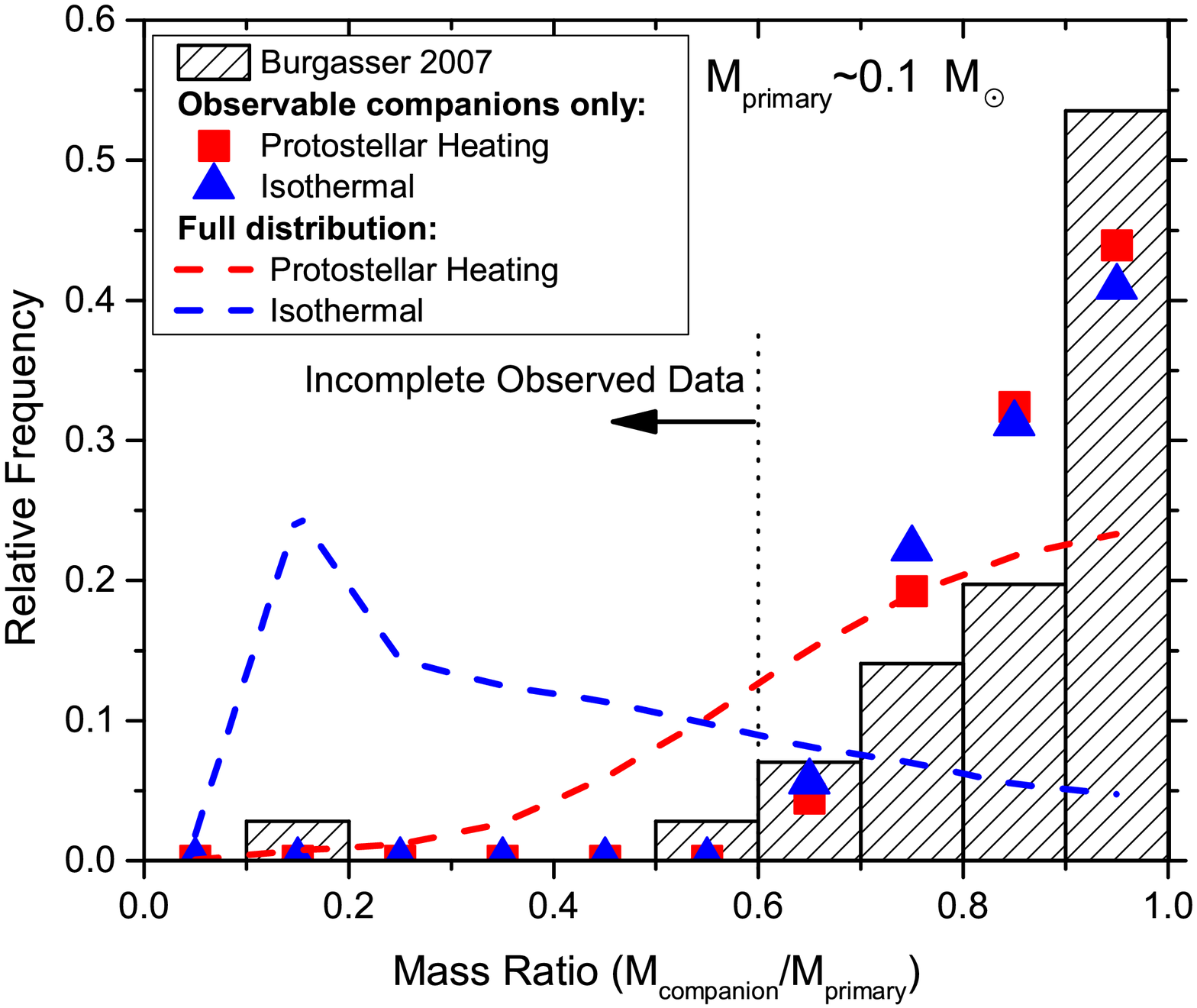}
\caption{The relative frequency of \textbf{most massive} companions of Solar type stars (left) and VLM stars (right). Red and blue points and dashed lines show the results of our simulations with and without applying the observational bias, as indicated in the legend, while hatched histograms show the observations of \citet{Raghavan2010} for Solar type stars and \citet{Burgasser2007} for VLM stars. While both the isothermal and radiative models are consistent with observations after applying completeness limits, the isothermal simulations predict the existence of a very large number of currently-unseen low-mass companions that would be detectable in deeper observations. Also, none of our results reproduce the peak at unit relative mass which could be the result of either preferential dynamical hardening/ejection or missing physics in the model (e.g. disk physics).}
\label{fig:companion_dist_hierarchical_observ}
\end {center}
\end{figure*}

One of the key observed properties of binaries is the apparent flat mass distribution of companion masses with a cutoff at very low masses (the so-called \myquote{brown dwarf desert}). In Fig. \ref{fig:companion_dist_hierarchical_observ} we test to what extent our models can reproduce this observation by comparing  the mass distribution of the \emph{most massive} companions in our simulations with observations of this quantity for Solar type and very low mass (VLM) stars ($M\sim 0.1\,\solarmass$). Although in principle we could compute other mass ratios (e.g., the mass ratios of all pairs of stars in multiple systems, c.f.~\citealt{Raghavan2010}), we focus on the most massive companions because these are the most robustly determined from observations. It is extremely challenging observationally to identify secondary and tertiary companions of a star in a triple or quadruple system. As a result, observations are most likely to discover the most massive companion rather than all companions, making the primary to secondary mass ratio the most well-determined. This also has the advantage that the most massive companion is the least likely to be ejected by dynamical processes. For our heated models, however, in practice it makes relatively little difference whether we include all companions or just the most massive one.

Examining Figure \ref{fig:companion_dist_hierarchical_observ}, it is clear that the isothermal and heated models are both roughly consistent with observations \emph{after} the observational bias is applied. The primary exception is that both models somewhat underpredict the frequency of near-equal mass companions; such companions can plausibly be attributed to disk fragmentation, which tends to produce mass ratios close to unity \citep{Bate2000}. It is important to note that without the observational bias the isothermal model predicts an overwhelming number of very low mass ratio companions. Meanwhile the results for Solar type stars in the heated case is only slightly affected by observational bias, which means that the brown dwarf desert is not an observational bias.

\begin{figure*}
\begin {center}
\includegraphics[width=0.475\linewidth]{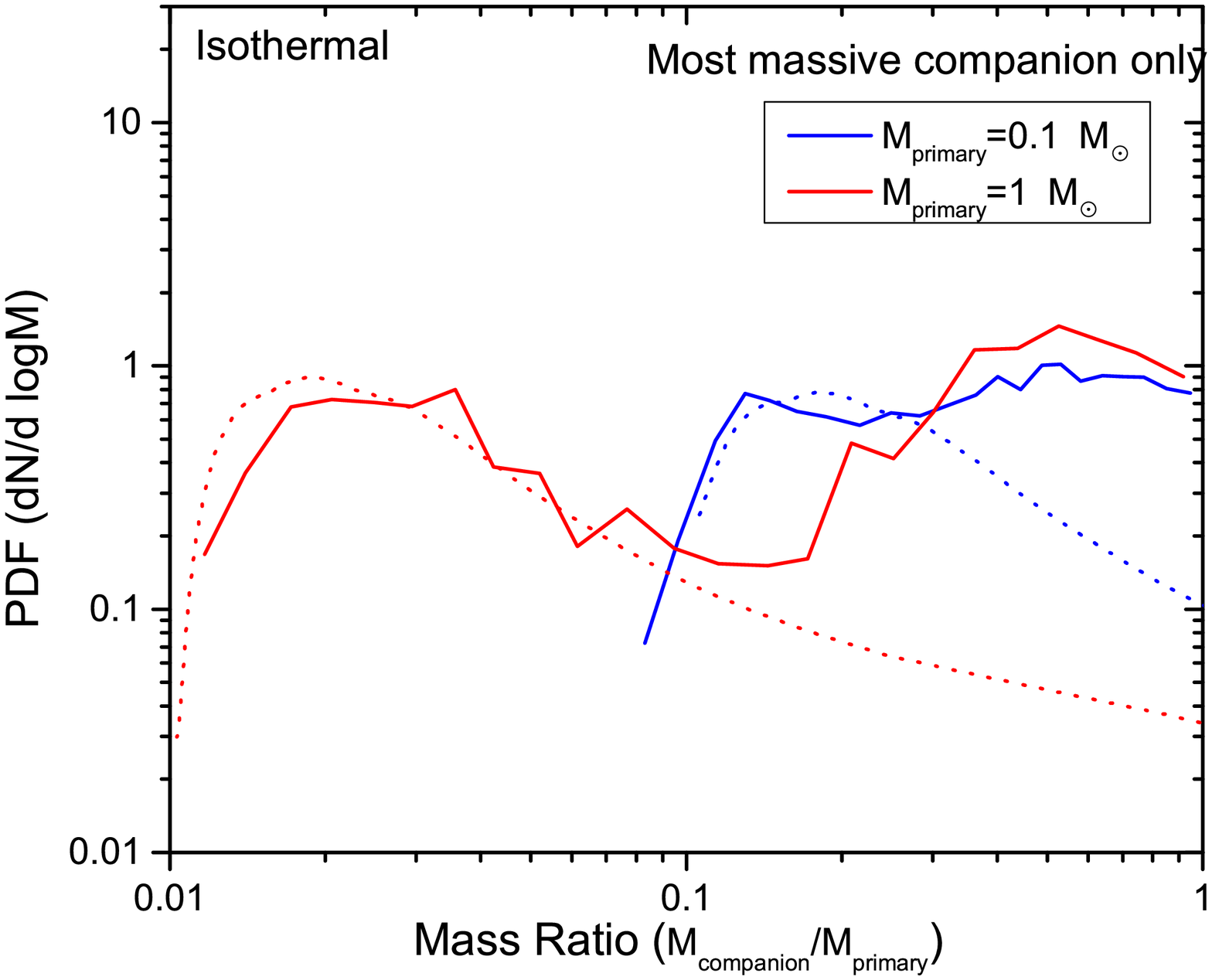}
\includegraphics[width=0.475\linewidth]{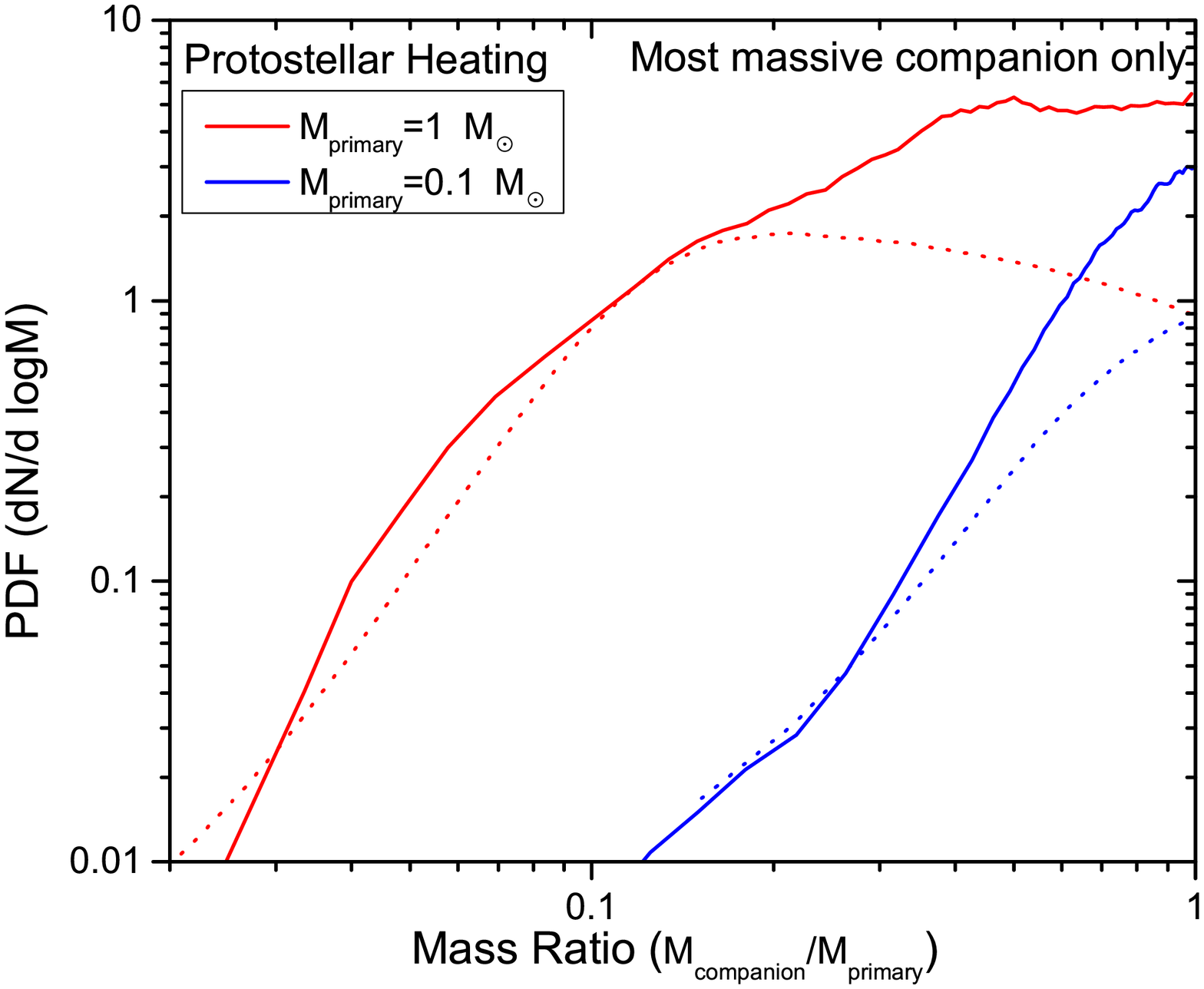}\\
\caption{
The two panels show the PDF of mass ratios between the two most massive stars in a given system, for primary masses $M_{\rm primary} = 0.1$ $\solarmass$ (solid blue) and $M_{\rm primary} = 1$ $\solarmass$ (solid red). We show these quantities both for the isothermal case and the model with protostellar heating, as indicated. For comparison, the dotted lines show the distributions that would result from the null hypothesis that companions are randomly drawn from the IMF. Note that the underlying simulation results shown here are identical to those shown in Figure \ref{fig:companion_dist_hierarchical_observ}, but they have been binned differently here for clarity. }
\label{fig:companion_dist_hierarchical_massive}
\end {center}
\end{figure*}

To gain further insight into why our heated models are able to reproduce the brown dwarf desert, while our isothermal models fail, let us compare these companion mass distributions with the null hypothesis that that companion masses are randomly drawn from the IMF\footnote{Binaries forming from randomly sampling the IMF has been ruled out \citep{Reggiani_Meyer_2011}, making it an important test for theoretical models.}. Fig. \ref{fig:companion_dist_hierarchical_massive} compares our measured companion mass ratio distribution with that we would expect under the null hypothesis, again considering only the most massive pair in a given star system\footnote{Computing the null hypothesis distribution requires some care, because for systems with $>2$ stars, even if all companions are drawn randomly from the IMF, the mass distribution for the most massive companion does not follow the IMF. Specifically, suppose we have an IMF $dN/dM = p(M)$, so that the cumulative distribution function (CDF) of masses (i.e., the probability that a randomly chosen star has mass $<M$) is $P(M) = \int_0^M p(M)\, dM$. Now consider a system where the primary has mass $M_p$. Since we require companion masses to be smaller than $M_p$, they follow the conditional CDF $P(M\mid M_p) \propto \int_0^{\min\left(M,M_p\right)} p(M) \, dM$, which for $M < M_p$ has the same shape as the CDF for single stars. However, now consider a system consisting of $n$ stars. The most massive companion has a mass $<M$ only if all $n-1$ companions have mass $<M$, and if the companion masses are independent the probability of this is $P(M \mid M_p)^{n-1}$. This does not have the same shape as the single star CDF. For the purposes of Figure \ref{fig:companion_dist_hierarchical_massive}, we account for this effect by generating our null hypothesis lines as a weighted sum $P_1(M\mid M_p) = \sum_{n>1} w_n(M_p) P(M\mid M_p)^{n-1}$, where both the single star CDFs $P(M\mid M_p)$ and the relative frequencies $w_n(M_p)$ of multiplicity $n$ are measured directly from the simulations for each primary mass $M_p$.}. The figure shows that in no case are the results consistent with random sampling of companions from the IMF. In the isothermal case the companion distribution for both Solar and VLM primaries follows the IMF for very low mass companions, but that there is a significant excess of companions at mass ratios $\sim 0.5 - 1$. In the heated case the situation is qualitatively similar, in that mass ratios near unity are overrepresented compared to the null hypothesis.

Now let us consider the implications of this finding for the brown dwarf desert. The companion mass ratio distribution is a product of two factors: the underlying IMF of all stars, and any biases imposed by the fact that the stars whose mass distribution we are computing are non-primaries. With or without heating, we find that mass ratios near unity are favoured compared to a null hypothesis of random IMF sampling. That is, if we collect two samples of stars with the same upper mass limit, and for one sample we randomly select only non-primary stars and for the other we randomly select stars without regard to multiplicity characteristics, the non-primary sample will typically be more massive. For Solar-type primaries, the combination of a bias towards higher mass companions and the overall negative slope of the IMF near 1 $\solarmass$ (so that lower mass stars are more probable overall) yields a relatively flat mass ratio distribution -- the IMF shape and the bias nearly cancel.

Now let us consider VLM stars. For VLM primaries, the bias towards equal mass companions is qualitatively similar to that for Solar-type stars. For our isothermal case, and unlike in reality, the IMF slope near $0.1$ $\solarmass$ is also about the same as that near $1$ $M_\odot$, due to the overall scale-free nature of isothermal fragmentation. Because both the IMF slope and the bias are about the same for Solar and VLM stars, the distribution of companion mass ratios is also qualitatively similar. For our heated case, as in reality, we have a very different situation. The slope of the IMF is negative near $1$ $\solarmass$, but positive (or at least close to flat) near $0.1$ $\solarmass$. As a result, for VLM primaries both the bias towards massive companions and the IMF itself favour more massive objects as companions. The result is a companion mass ratio distribution that is sharply biased towards stellar companions and away from brown dwarfs, producing the observed brown dwarf desert. We therefore find that the brown dwarf desert is a result of the change in the IMF slope between $\sim 0.1$ and $\sim 1$ $\solarmass$, which in turn is imposed by thermal feedback causing a deviation from scale-free behaviour during gas collapse and fragmentation.

\subsubsection{Binary Separations}

\begin{figure}
\begin {center}
\includegraphics[width=\linewidth]{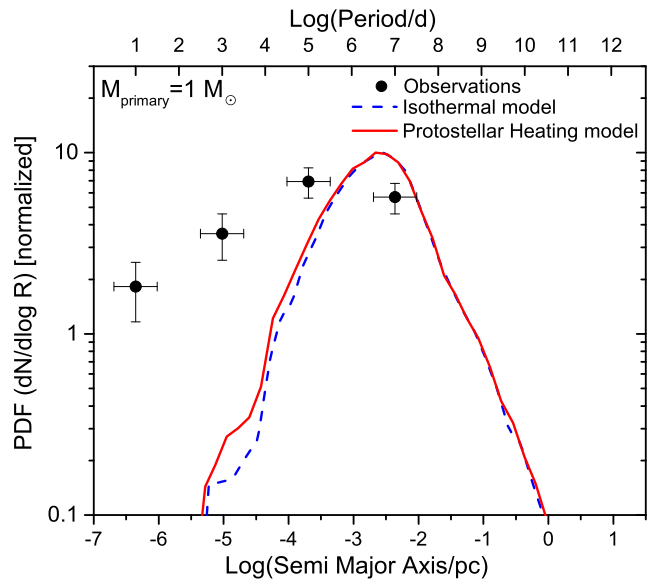}
\caption{Semi-major axis distribution for binaries with fixed primary mass ($\sim 1\,\solarmass$) in case of the isothermal model (dashed) and the case with protostellar heating (solid). The figure also includes the observed present day distribution from \protect\cite{Moe_DiStefano2016} for solar type stars. The observed period distribution of solar type stars could plausibly be explained by common core fragmentation at large scales, but there a serious discrepancy for short-range binaries, further implying that protostellar disk fragmentation or dynamical effects play a crucial role.}
\label{fig:semi_major_distrib}
\end {center}
\end{figure}

In addition to the mass ratio distribution, our spatially-resolved model allows us to examine the predicted semi-major axis distribution of binaries. We do so in Fig. \ref{fig:semi_major_distrib} for Solar-type stars. The distribution appears peaked which comes from the peak of the companion mass ratio distribution (Fig. \ref{fig:companion_dist_hierarchical_massive}) with the  corresponding length scale of $\sim\frac{G M_p}{c_s^2}\approx 0.05\,\pc$.

Comparing with the observations from \cite{Marks2011} we can see that on large scales our model of common core fragmentation seems to very roughly reproduce the present day observations. Although our results only give the \myquote{at-formation} period distribution, the comparison is still meaningful because, as explained in Sec. \ref{sec:method}, we have attempted to limit the systems we count in our model to hierarchical systems that should be dynamically stable. In any case, it is clear that, similar to the case of the 2D correlation function (Fig. \ref{fig:observations}), turbulent fragmentation is unable to reproduce the observations on small scales. Another source for such binaries is required at $\leq 100\,\rm{AU}$, for which protostellar disk fragmentation is a good candidate\footnote{
It should be noted that disc fragmentation simulations also fail to produce extremely close binaries ($\leq 10\,\rm{AU}$). These are likely to have either formed from a wider binary whose separation decreased due to orbital decay (e.g. \citealt{Stahler_2010,Korntreff_orbital_period}), or from exchange interactions in star clusters.}. Note that decreasing $R_{\rm min}/R_0$ would technically improve the fit (see Appendix \ref{sec:numerical_error}), but this would require unphysically low values.

\section{Conclusions}\label{sec:conclusions}

The aim of this paper is to investigate the origin of the stellar correlation function and multiplicity statistics, and in particular to understand which features of these distributions result from pure scale-free isothermal fragmentation, and which bear the imprints of scale-dependent stellar feedback. Using the MISFIT semi-analytical turbulent fragmentation framework of \framework~and \frameworkrad we find that the shape of the correlation function is almost entirely set by isothermal turbulence. Stellar feedback, which operates primarily on small scales, has little effect. On smaller scales ($\leq 100\,\rm{AU}$) both a purely isothermal model and one including stellar radiation feedback underpredict the stellar correlation, suggesting that our turbulent fragmentation models lack certain small scale physics (likely protostellar disk fragmentation). As with the correlation function, we find that our models provide a reasonable match to the observed the binary period distribution at large separations regardless of whether we include protostellar heating or not, but that we fail to produce enough very close binaries. We again conjecture that these close binaries are a result of disk fragmentation and N-body interaction, which our model does not include.

The situation for the mass ratios and multiplicity fraction of binaries is quite different. Isothermal fragmentation produces far too many multiple stars compared to what is observed, with even $\sim 1$ $\solarmass$ stars predicted to have multiplicities near unity. Adding protostellar heating substantially improves the situation, though the multiplicity fraction is still somewhat too high, likely because our models do not include dynamical evolution that will disrupt unstable systems. These differences, however, are almost completely washed out by the observational biases. 

Most interestingly, if we neglect the observational bias we find that while turbulent fragmentation with or without protostellar heating can adequately reproduce the observed companion mass distribution for Solar type stars (except for very low mas companions), but only when protostellar heating is included can we reproduce the mass distribution of companions for low-mass primaries. In particular, only our models including radiative feedback reproduce the ``brown dwarf desert'', whereby the companions to low mass stars ($\sim 0.1$ $\solarmass$) are overwhelmingly stellar objects (i.e., close to a mass ratio of unity) rather than brown dwarfs. Models that include only scale-free physics predict a companion mass ratio distribution for low mass stars that is qualitatively similar to that for Solar-type stars, a direct consequence of the scale-free nature of these models. In contrast, protostellar heating suppresses the number of brown dwarfs relative to stars, so that the companion mass ratio distribution is very different for Solar-type stars that lie above the IMF peak and low-mass stars that lie at ore below it. We therefore conclude that the brown dwarf desert is a consequence of the physical mass scale imprinted by protostellar heating into the otherwise scale free star formation process.

\acknowledgments
Support for PFH and DG was provided by an Alfred P. Sloan Research Fellowship, NASA ATP Grant NNX14AH35G, and NSF Collaborative Research Grant \#1411920 and CAREER grant \#1455342. MRK is supported by NASA ATP grant NNX15AT06G and Australian Research Council grant DP160100695. Numerical calculations were run on the Caltech compute cluster ``Zwicky'' (NSF MRI award \#PHY-0960291) and allocation TG-AST130039 granted by the Extreme Science and Engineering Discovery Environment (XSEDE) supported by the NSF. \\

\bibliographystyle{mnras}
\bibliography{bibliography}

\appendix

\section{Improvements to Previous Model}\label{sec:corrections}

Two papers (\citealt{TurbFramework, guszejnov_feedback_necessity}) have been published so far using the MISFIT (Minimalist Star Formation Including Turbulence) semi-analytical star formation framework as this paper. Since the publication of those results, several improvements have been made to the algorithm, all of which are implemented for this paper. These do not change any of the published qualitative IMF results (e.g. general shape, sensitivity to initial conditions). They include:
\begin{itemize}
	\item Correction of a bug that suppressed fragmentation at the end of the cloud evolution, violating self-similarity at a weak level. The effects on our previous work are small, but are substantial on the statistics on low mass companions. This is now fixed. 
	\item Fragments are properly tracked and taken into account for the evolution of their parent (e.g. their contribution to the gravitational potential is taken into account as long as the parent has not yet contracted beyond their position). This causes no qualitative difference.
	\item Instead of using an absolute termination scale (taken to be $R_{\rm min}=10^{-4}\,\pc$ in the previous papers, roughly the size of protostellar disks), the collapse of clouds now terminates once clouds have contracted to a fixed fraction of their initial radius, chosen to be roughly when angular momentum support becomes dominant. This assumes that the source of angular momentum for clouds is from random turbulent motion. The resulting distribution for $\beta=\frac{E_{\rm rot}}{E_{\rm pot}}$ is strongly peaked around a few percent (\citealt{Burkert_Bodenheimer_rotation_2000}). If collapse happens at constant virial parameter than the size scale where angular momentum starts dominating is $\beta R_0$. This is the scale where the cloud flattens and forms a disk, which we choose as our termination point.\footnote{The choice of a relative termination scale instead of an absolute value has the added benefit of imprinting no absolute length scale into the problem, preserving self-similarity.}
	\item We set a lower limit of $0.007\,\solarmass$ on fragment masses based on the opacity arguments of \cite{LowLyndenBell1976}. This is in fact equivalent to a simplified EOS model, where we terminate the fragmentation once the cloud reaches the adiabatic limit. This provides a natural termination for the fragmentation cascade in our \myquote{isothermal} models, otherwise our results would not converge (see \frameworkrad)
\end{itemize}

\section{Comparison with detailed hydrodynamic simulations}\label{sec:Bate}

There have been a number of hydrodynamical simulations attempting to find the multiplicity statistics and separation distribution of newly formed stars (e.g. \citealt{Goodwin04a,Delgado_Donate_2004,Bate09a,Bate12a,Offner2010,Krumholz12a}). The semi-analytical approach we present in this paper has several advantages over these (e.g. faster, no absolute resolution limit, starts from GMC) but at the cost of several strong assumptions, so it is crucial that we compare our results with theirs. We choose to compare with the simulations of \cite{Bate12a}, as these have the largest sample of multiple systems. Since we know the full binary distribution from the simulations, we make this comparison without the observational completeness correction that we apply when comparing to observations in the main text. 

\begin{figure}
\begin {center}
\includegraphics[width=\linewidth]{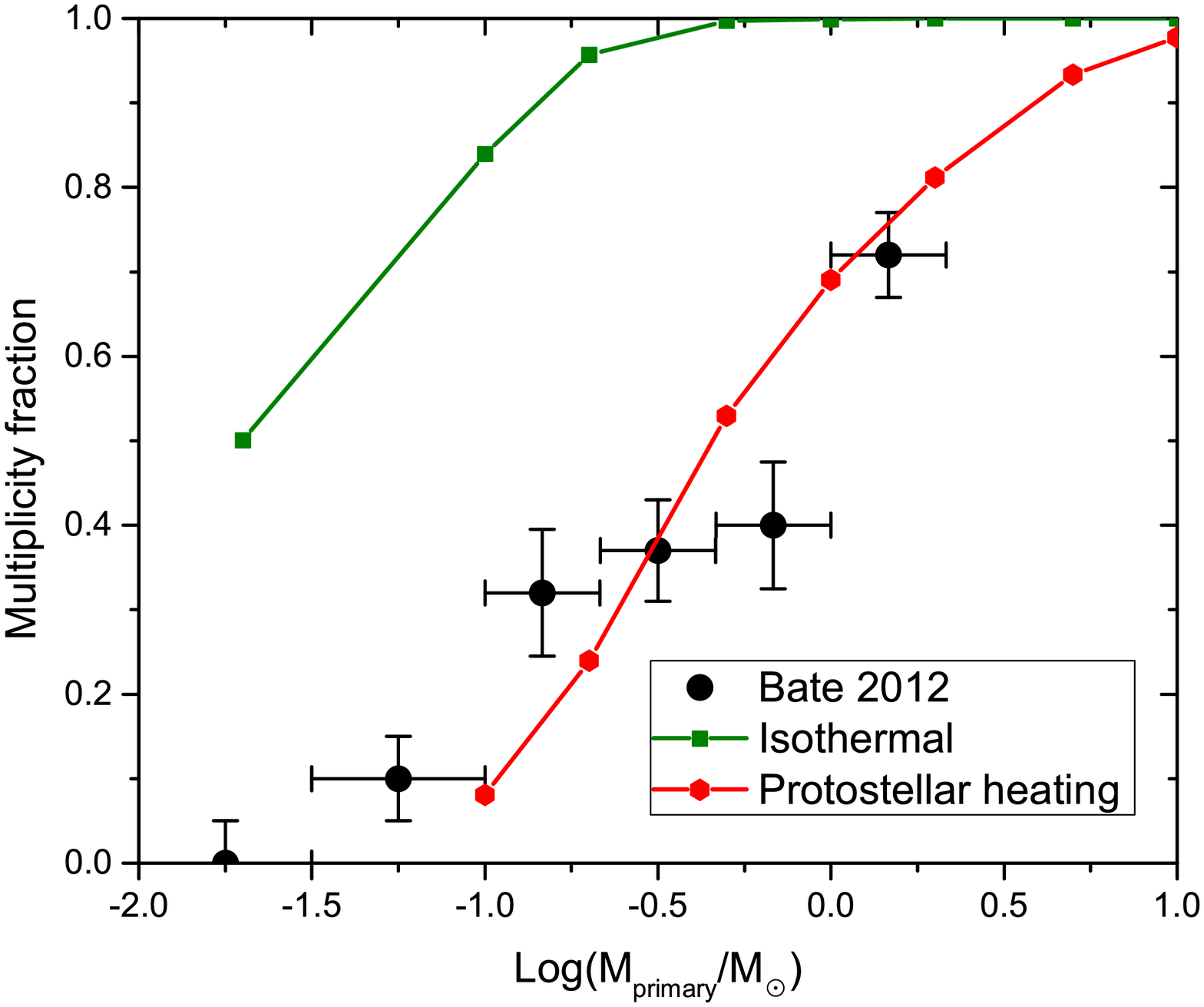}
\caption{
The multiplicity of stars of different masses in the isothermal case (\textit{Isothermal - MW}) and the model with protostellar heating (\textit{Heating - MW}) compared to the results of \protect\cite{Bate12a} (black circles with error bars).}
\label{fig:multiplicity_fraction_Bate}
\end {center}
\end{figure}

\begin{figure*}
\begin {center}
\includegraphics[width=0.475\linewidth]{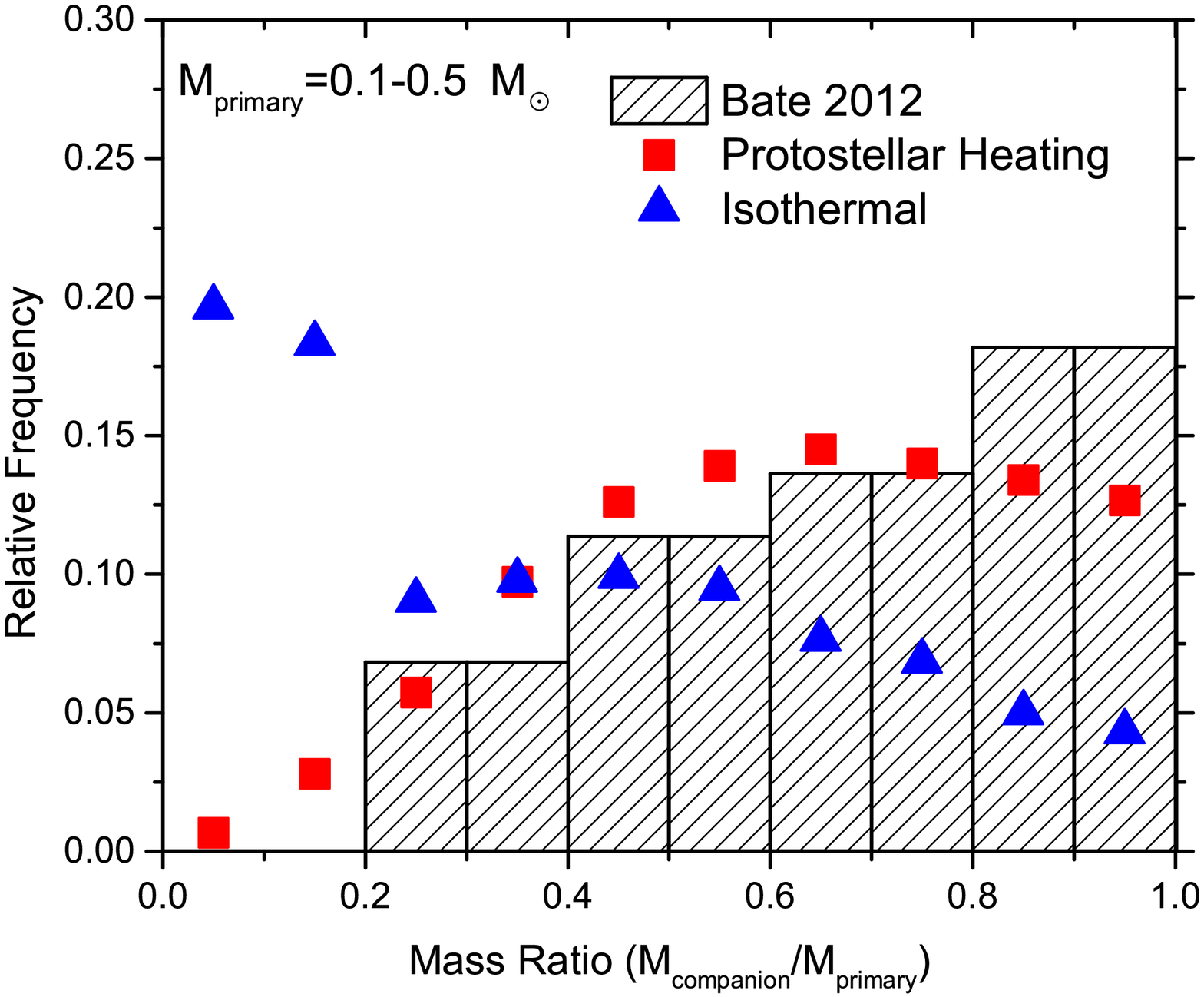}
\includegraphics[width=0.475\linewidth]{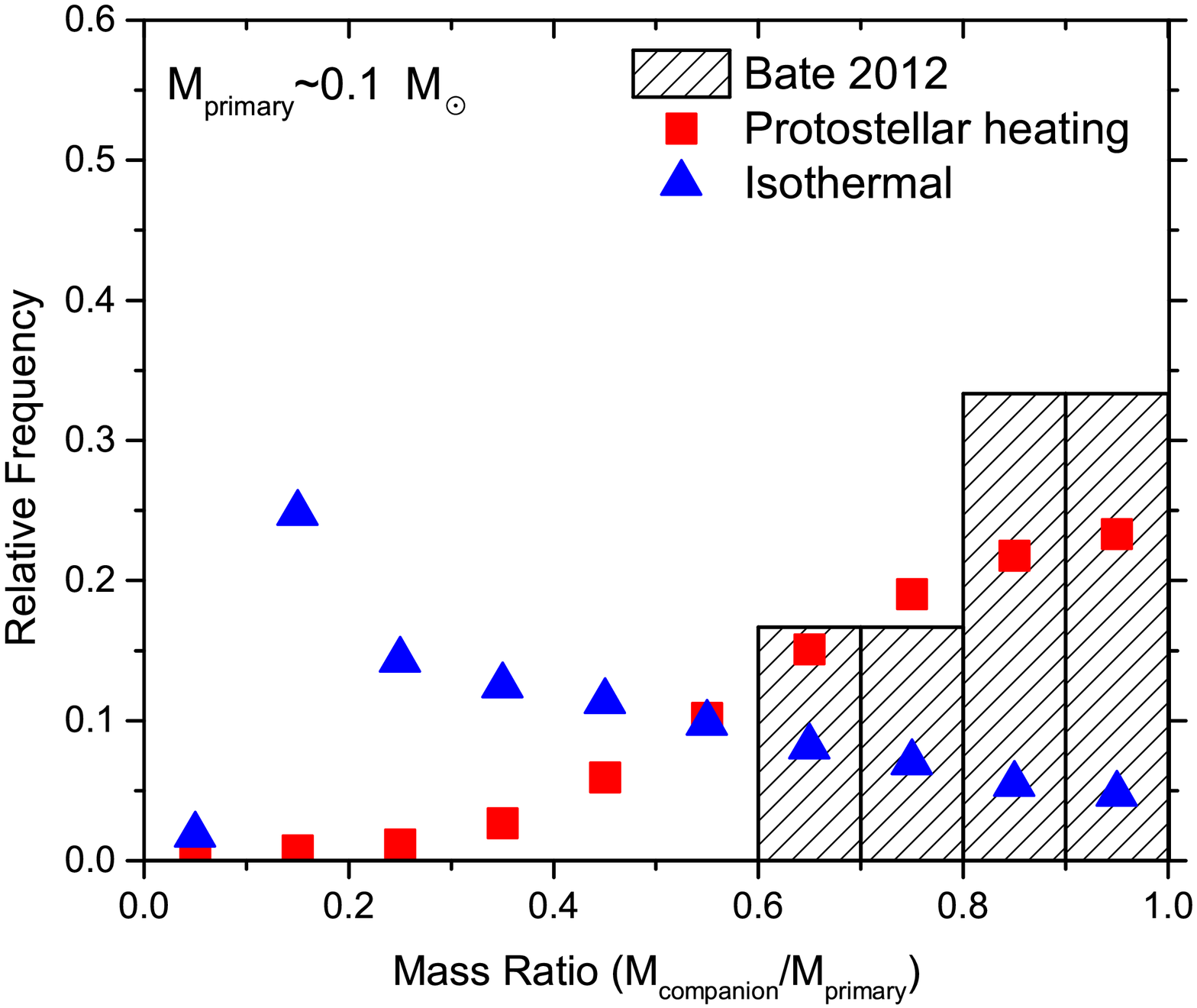}
\caption{The relative frequency of \textbf{most massive} companions of small stars (left, $M_{\rm primary}=0.1-0.5\,\solarmass$) and VLM stars (right, $M_{\rm primary}=0.1\,\solarmass$). Red and blue points show the results of our simulations without applying the observational bias, while hatched histograms show the results of \protect\cite{Bate12a}.}
\label{fig:companion_dist_hierarchical_Bate}
\end {center}
\end{figure*}

\begin{figure}
\begin {center}
\includegraphics[width=\linewidth]{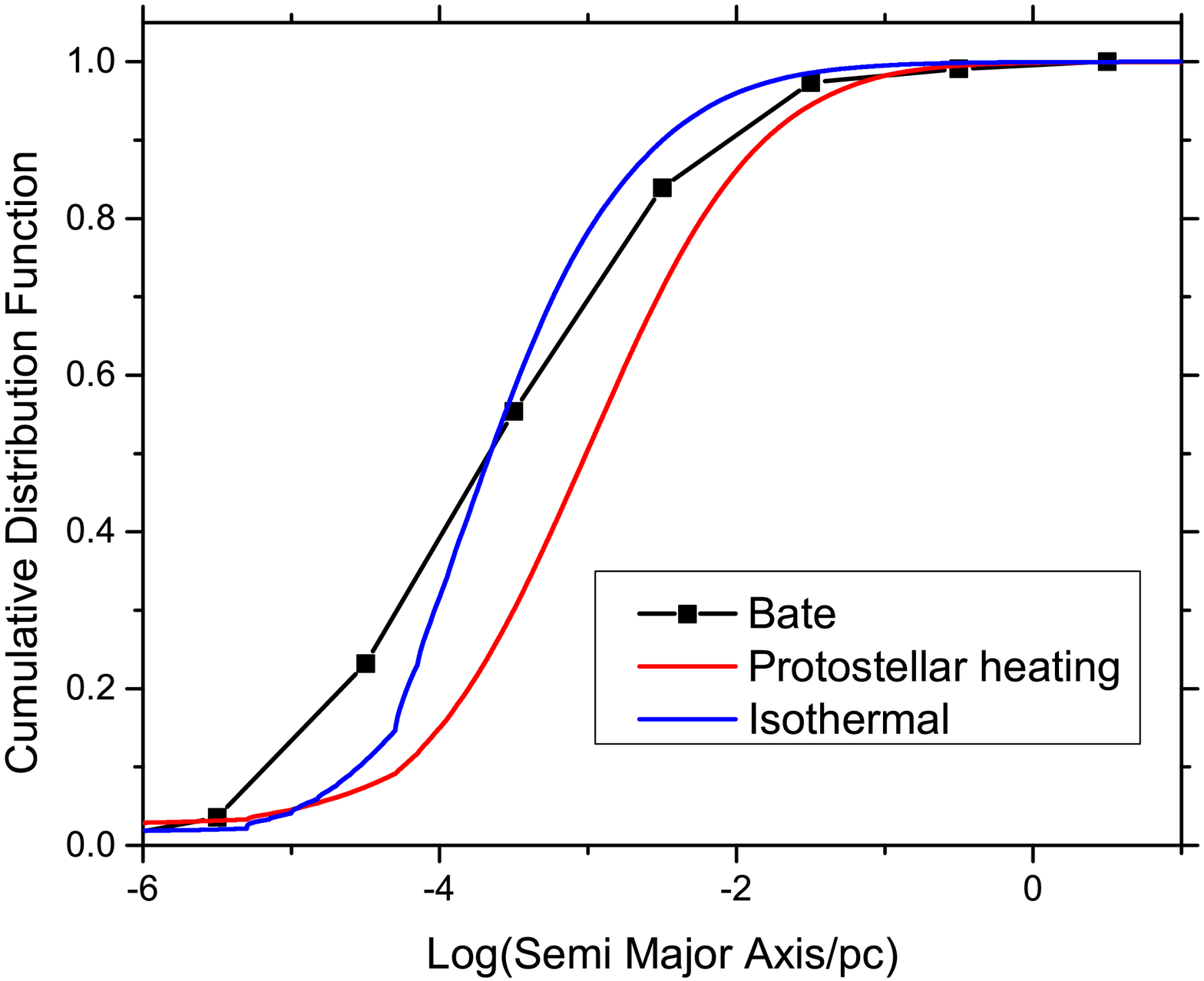}
\caption{Cumulative Semi-major axis distribution for multiple systems where ($M_{\rm primary}> 0.1\,\solarmass$) in case of the isothermal model (blue) and the case with protostellar heating (red) along with the results of \protect\cite{Bate12a}. This includes \emph{all} separations, hierarchical triples and quadruples contribute 2 and 3 values respectively.}
\label{fig:semi_major_distrib_Bate}
\end {center}
\end{figure}

Figures \ref{fig:multiplicity_fraction_Bate}-\ref{fig:semi_major_distrib_Bate} show that our results are qualitatively, and in many cases quantitatively, consistent with the simulations of \cite{Bate12a}. An important difference between our current model and the traditional simulations is that MISFIT does not have a finite resolution limit, but it does neglect disk physics. This leads to the discrepancy at small separations shown in Fig. \ref{fig:semi_major_distrib_Bate}.

\section{Numerical Tests and Convergence}\label{sec:numerical_error}

In this section we show how the GMC mass$M_{\rm GMC}$, the termination scale $R_0/R_{\rm min}$, and the resolution parameter $N$ affect our results. To explore these questions, we have repeated our fiducial \textit{Heating - MW} run with different resolutions ($N=16,32,64)$, GMC masses ($M_{\rm GMC}=10^4,10^5,10^6$ $\solarmass$), and different termination scales ($R_0/R_{\rm min} = 10^{-3}, 10^{-2}, 10^{-1}$). The full list of runs performed in given in Table \ref{tab:simparam}. Note that unlike the results in the main text these are not modified to account for observational biases.

\begin{figure*}
\begin {center}
\includegraphics[width=0.45 \linewidth]{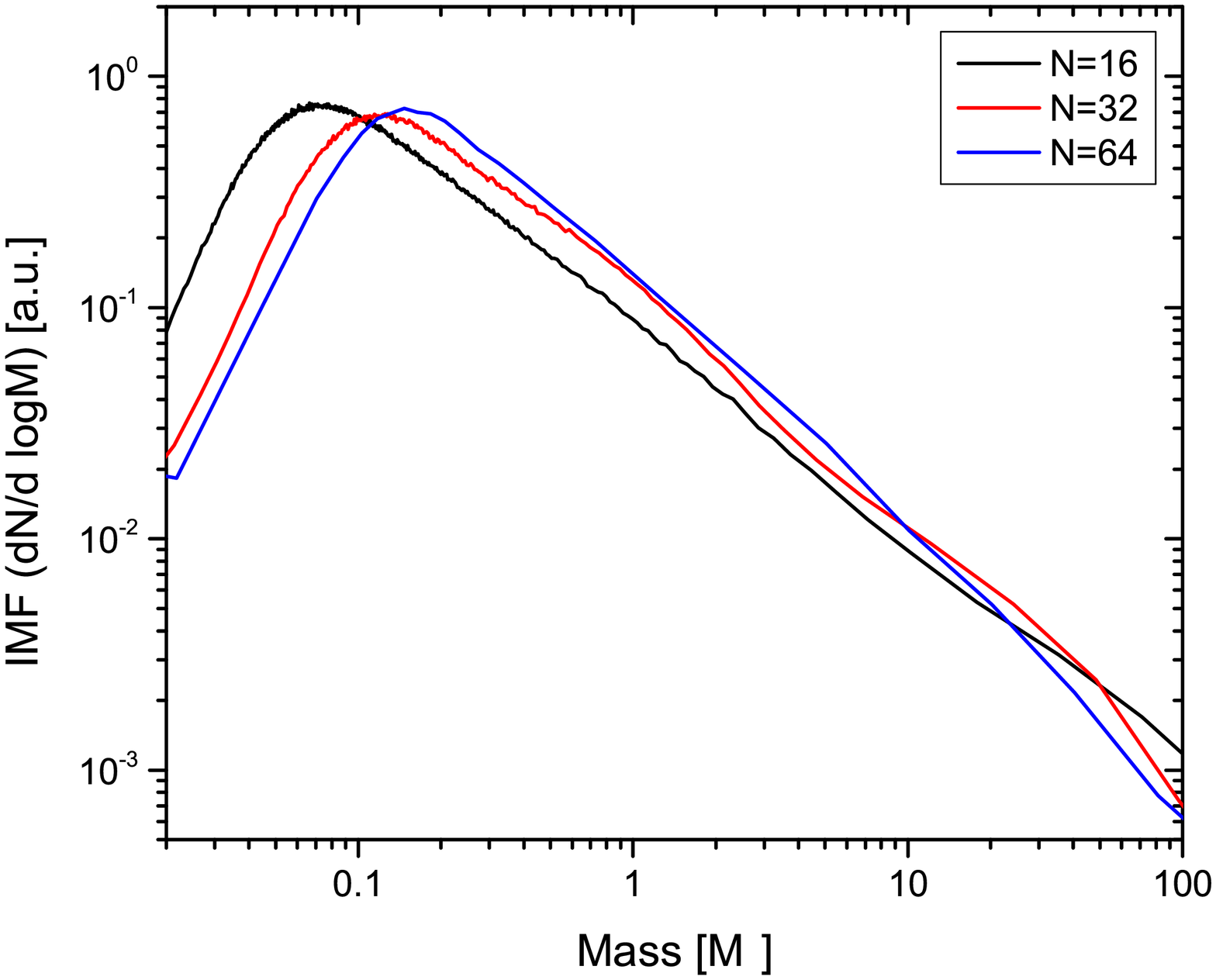}
\includegraphics[width=0.45 \linewidth]{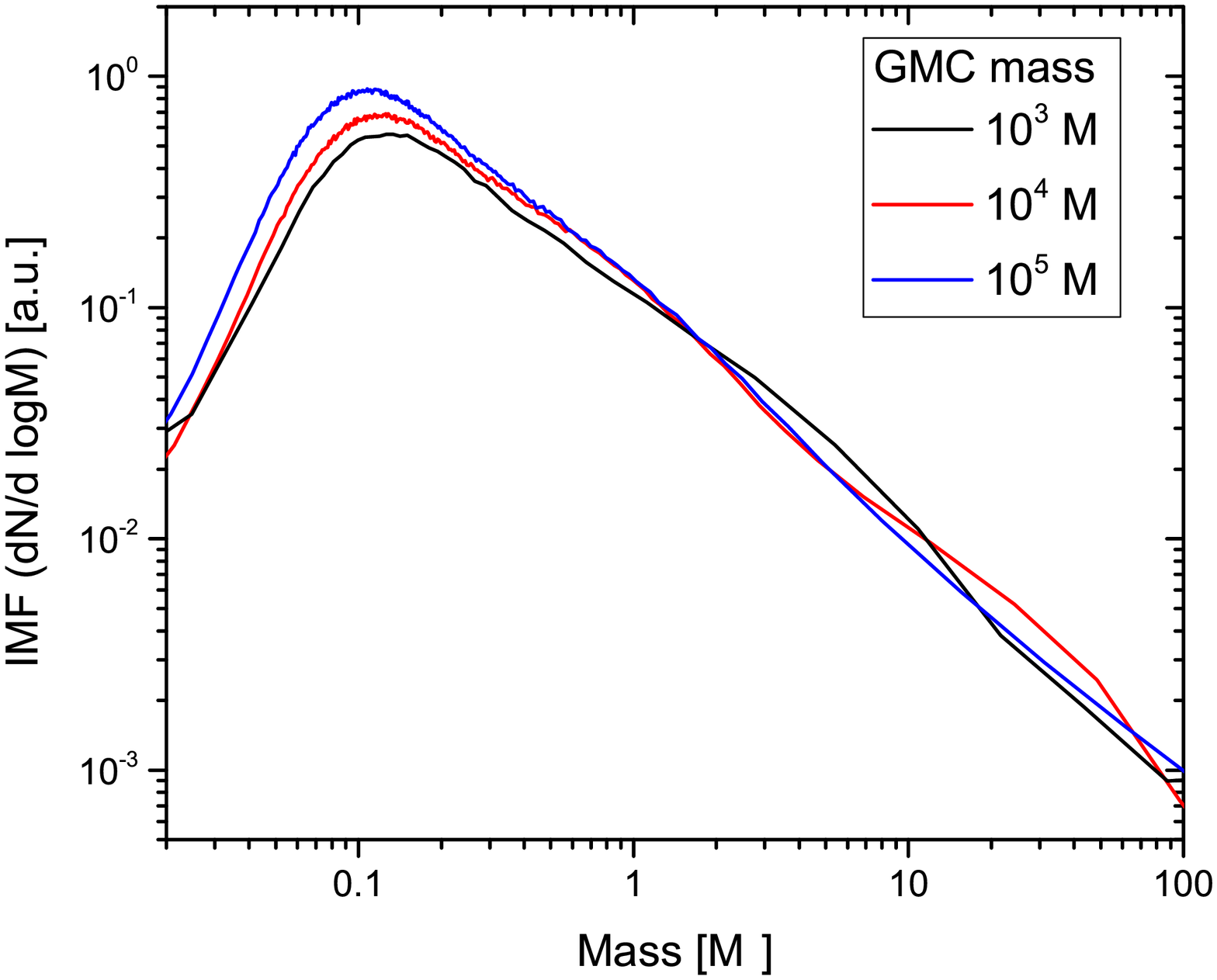}\\
\includegraphics[width=0.45 \linewidth]{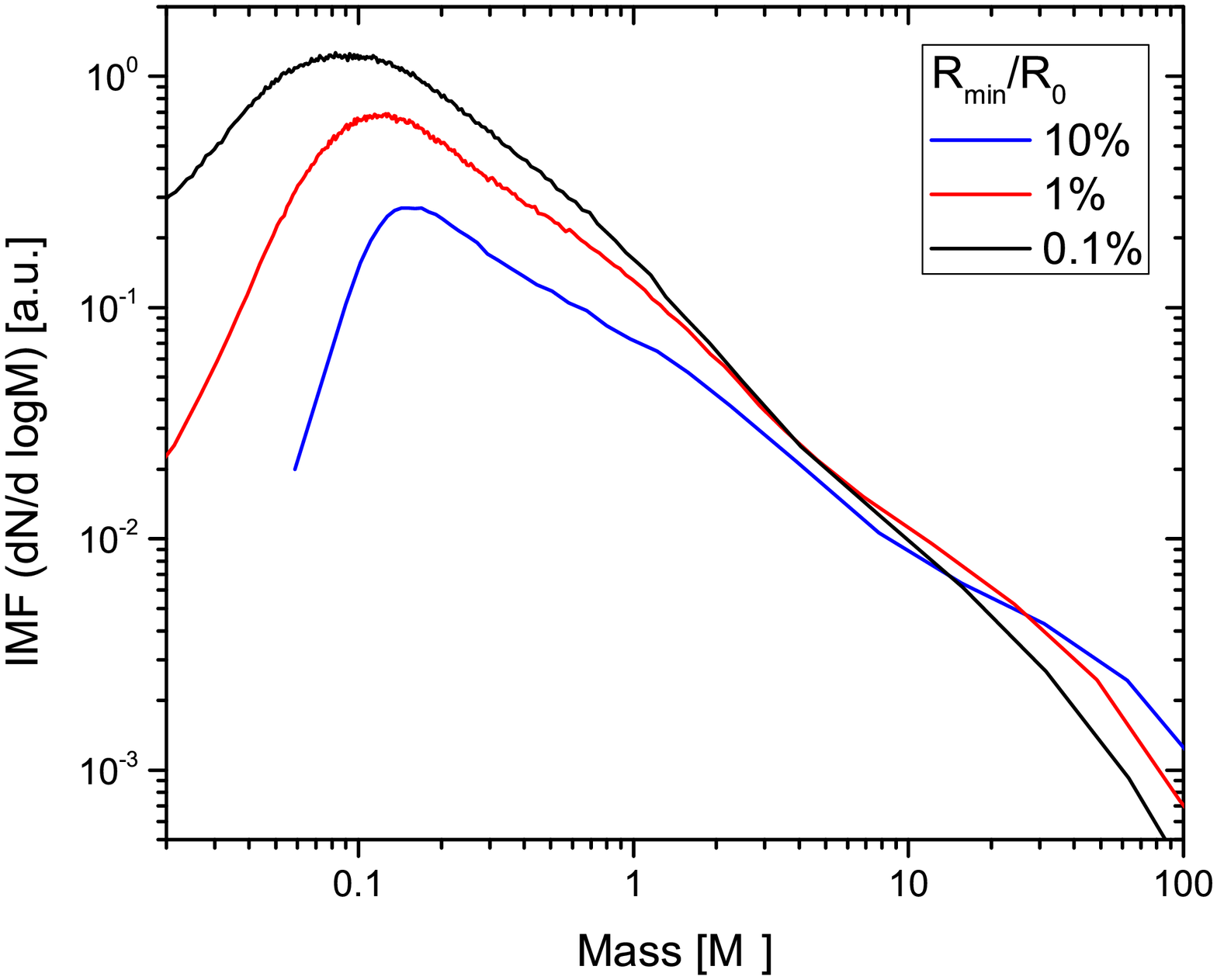}
\caption{Effects of the simulation resolution $N$ (top left), parent GMC mass (top right) and the relative termination scale $R_{\rm min}/R_0$ (bottom) on the IMF.}
\label{fig:IMF_effects}
\end {center}
\end{figure*}

We first examine how our results affect the same of the IMF produced by our models. Fig. \ref{fig:IMF_effects} shows that the shape of the IMF is robust to changes in $N$, reaching convergence around $N=32$. The location of the IMF peak and the high mass slope are also essentially insensitive to the GMC mass; the location of the peak does shift by an extremely small amount as we vary the GMC mass, as a result of its dependence on the Mach number of the turbulence; the two are related thanks to our assumption that clouds have virial ratios $\alpha_{\rm vir} \approx 1$. However this shift is only $\sim 10\%$ over a plausible range of GMC masses. The IMF shows its greatest sensitivity to the relative termination scale $R_{\rm min}/R_0$, particularly the abundance of brown dwarfs beyond the peak. In reality our choice of a single $R_{\rm min}/R_0$ value is an oversimplification, since real turbulence fields produce a distribution of rotational kinetic energies $\beta$, and thus a distribution of $R_{\rm min}/R_0$ parameters; the real IMF should therefore resemble a weighted average of the curves shown in Fig. \ref{fig:IMF_effects}.

\begin{figure*}
\begin {center}
\includegraphics[width=0.45 \linewidth]{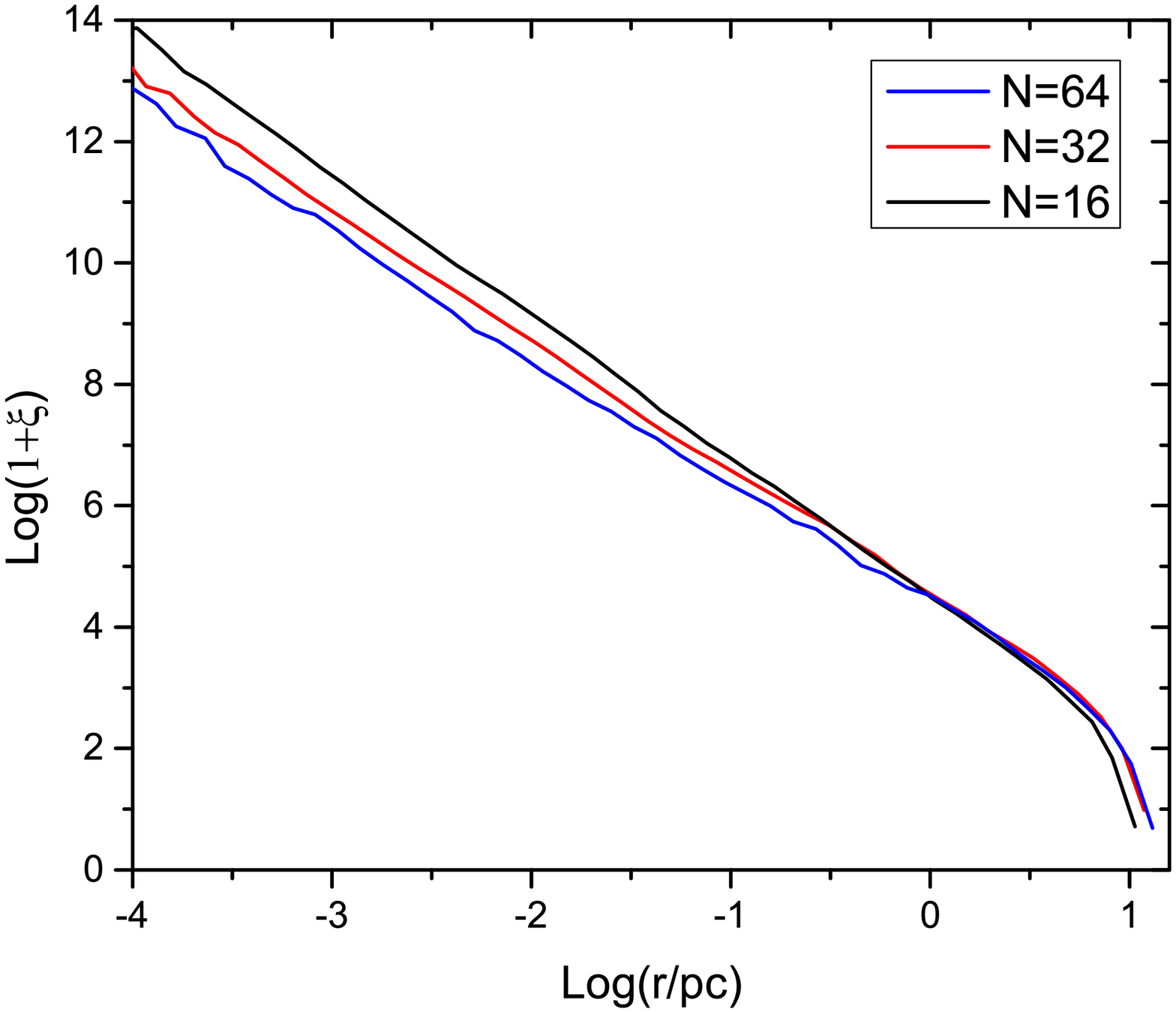}
\includegraphics[width=0.45 \linewidth]{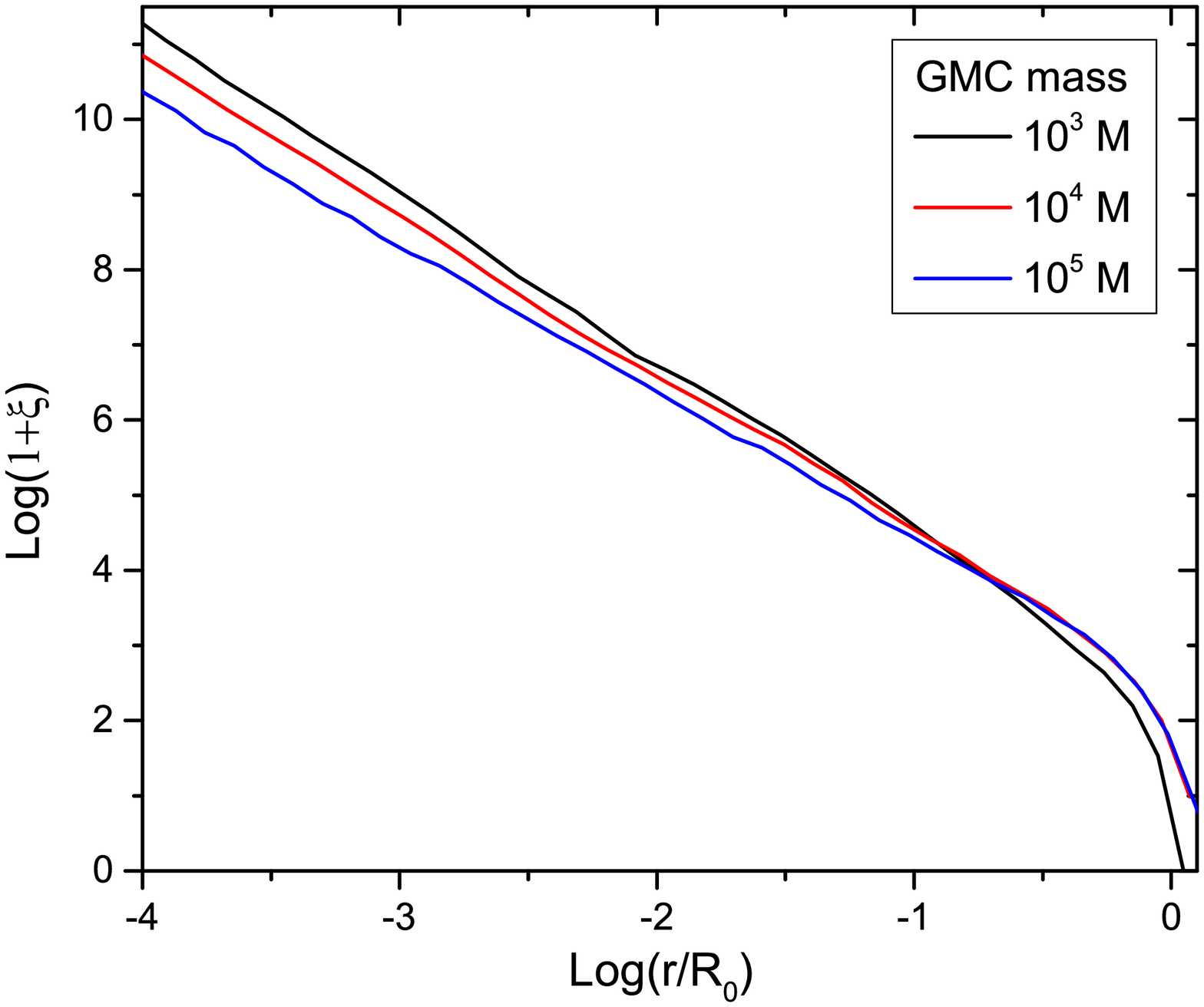}\\
\includegraphics[width=0.45 \linewidth]{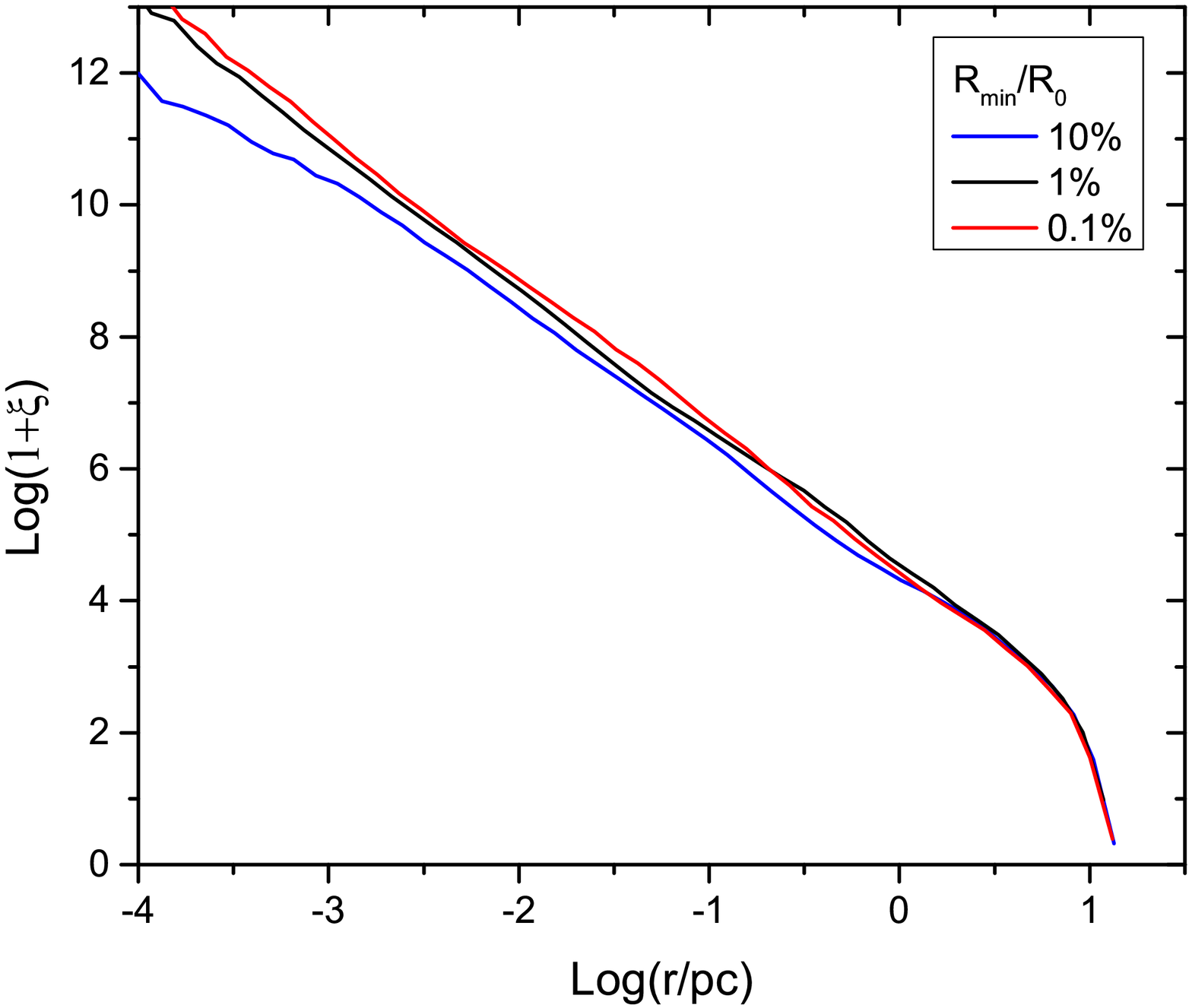}
\caption{Effects of the simulation resolution $N$ (top left), parent GMC mass (top right) and the relative termination scale $R_{\rm min}/R_0$ (bottom) on the stellar correlation function.}
\label{fig:corr_effects}
\end {center}
\end{figure*}

Fig. \ref{fig:corr_effects} shows how variation of our three parameters affects the stellar correlation function. As with the IMF, we find that the correlation function is insensitive to changes in both the resolution parameter $N$ and the initial GMC mass -- the former produces no noticeable differences past $N=32$, while the latter mostly rescales the outer cutoff/size scale. Also, as with the IMF, larger initial masses lead to verly slightly shallower slopes.This is consistent with the discussion in Section \ref{sec:corr_results} and Appendix \ref{sec:cantor}: larger masses mean stronger initial turbulence which in turn means easier fragmentation. However, as with the IMF, the effect is extremely modest. Finally, the bottom panel of Fig.\ref{fig:corr_effects} shows that the relative termination scale has no effect on the correlation function apart from introducing a small-scale cutoff.

\begin{figure*}
\begin {center}
\includegraphics[width=0.45 \linewidth]{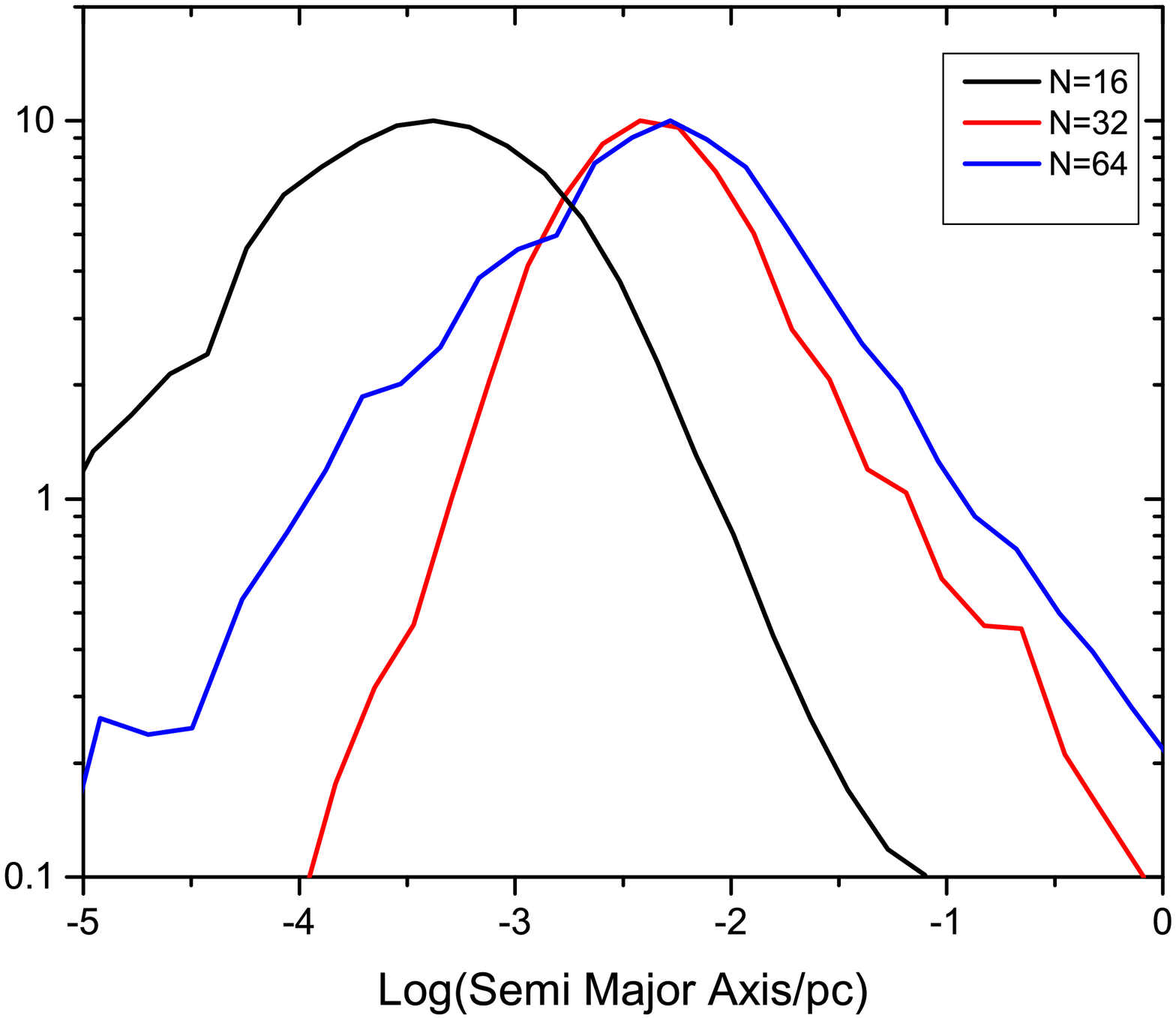}
\includegraphics[width=0.45 \linewidth]{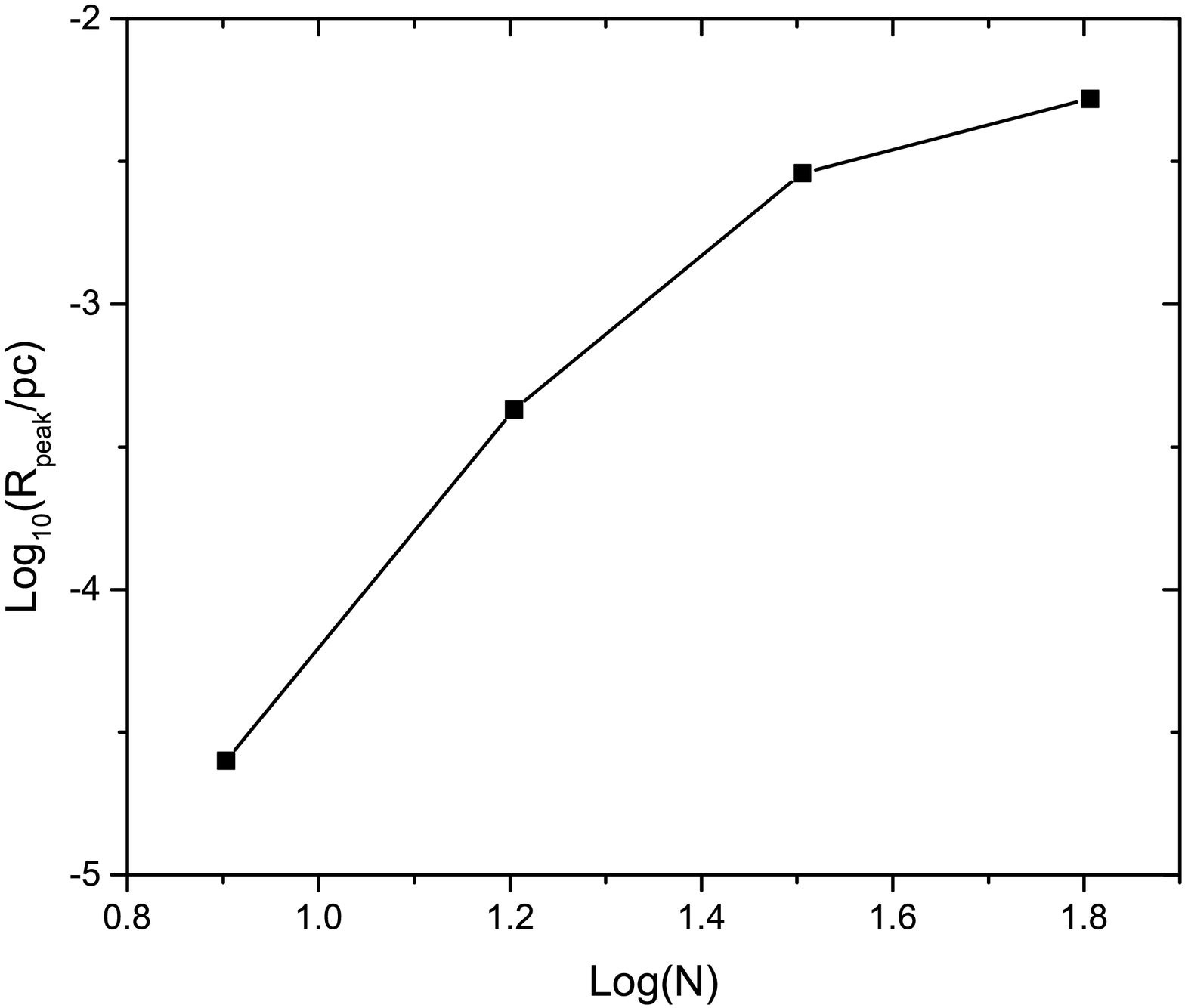}\\
\includegraphics[width=0.45 \linewidth] {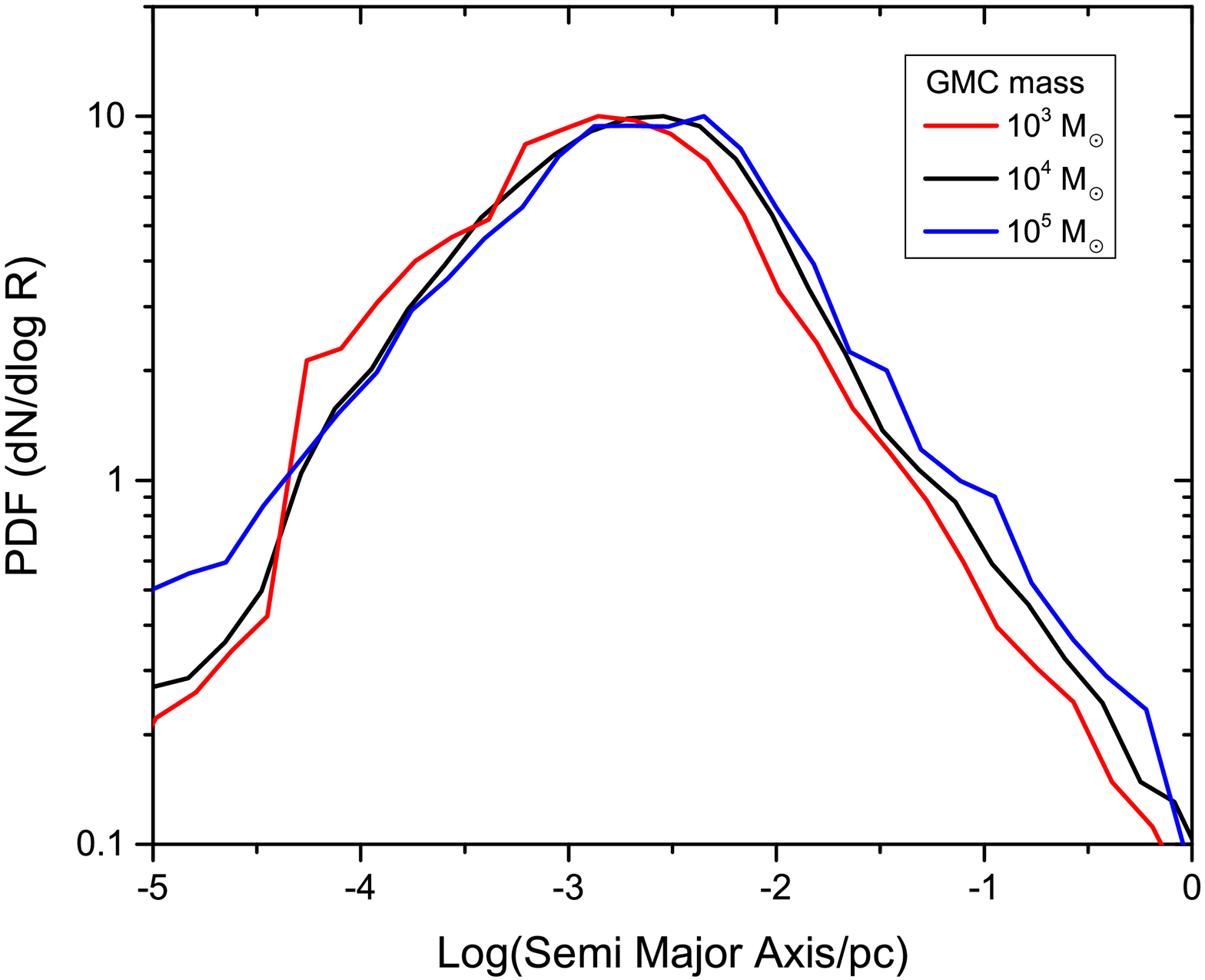}
\includegraphics[width=0.45 \linewidth] {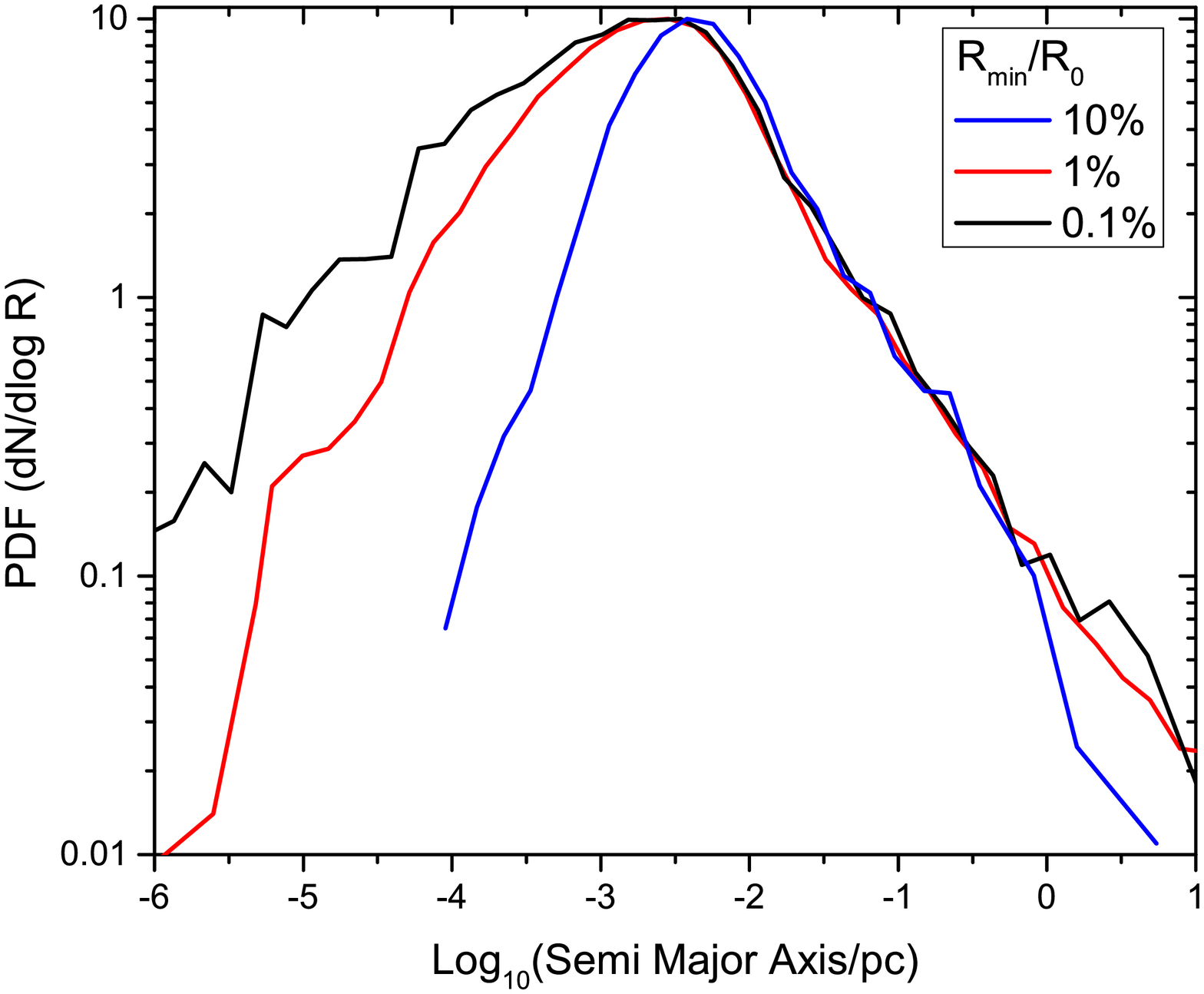}

\caption{Effects of the simulation resolution $N$ (top left and right), parent GMC mass (bottom left) and the relative termination scale $R_{\rm min}/R_0$ (bottom right) on the stellar correlation function.
\label{fig:semimajor_effects}}
\end {center}
\end{figure*}

 Fig. \ref{fig:semimajor_effects} shows the peak of the separation distribution converges above $N=32$, and the parent GMC mass has little effect on it. The relative termination scale $R_{\rm min}/R_0$ sets the width of the peak; its position is set by $\frac{G M_{\rm primary}}{c_s^2}$, as discussed in Sec. \ref{sec:multip_results}. Decreasing $R_{\rm min}/R_0$ increases the abundance of binaries at small separations, since it pushes the transition between common core fragmentation (which we are modeling) and disc fragmentation (which are are not) to smaller scales. However, as noted in the main text, only a value of $R_{\rm min}/R_0$ that is unphysically small would produce enough short-period binaries to be consistent with the observations.

\section{Cantor-like Model of Fragmentation}\label{sec:cantor}

\begin{figure}
\begin {center}
\includegraphics[width=\linewidth]{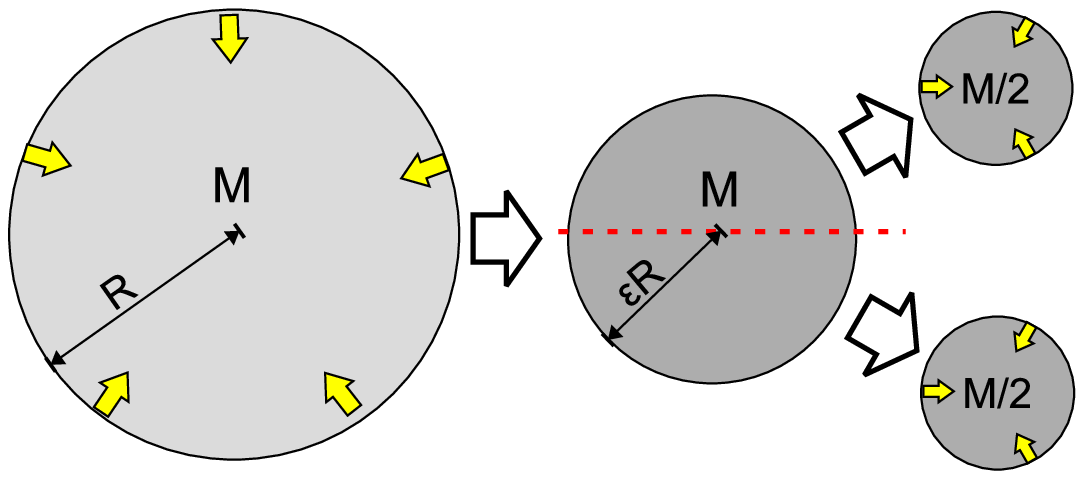}
\caption{The 3D Cantor-set-like toy model of isothermal fragmentation. Every cloud contracts to $\epsilon$ relative scale before breaking into two along a randomly chosen plane and the process repeats itself.}
\label{fig:cantor_model}
\end {center}
\end{figure}

One of the most important properties of isothermal fragmentation is that it is scale-free, so we expect self-similar, fractal-like structures to emerge. We can formulate a simple toy model to describe this process where self-gravitating clouds contract to about $\epsilon$ relative scale before breaking into two (see Fig. \ref{fig:cantor_model}) along a random axis. The distance of the two fragments is uniformly chosen between 0 and $R_{\rm parent}$. The fragments then rearrange themselves into spheres at the same density as their parent (meaning their radius is $2^{-1/3}R_{\rm parent}$). This model is very similar to the generalized Cantor dust (3D analogue of the generalized 1D Cantor set) that have the fractal dimensions of $D_{\rm set}=\frac{\ln 2}{\ln 2-\ln \epsilon}$ and $D_{\rm dust}=\frac{3 \ln 2}{\ln 2-\ln \epsilon}$, leading to a 3D correlation function of $r^{D-3}$.

\begin{figure}
\begin {center}
\includegraphics[width=\linewidth]{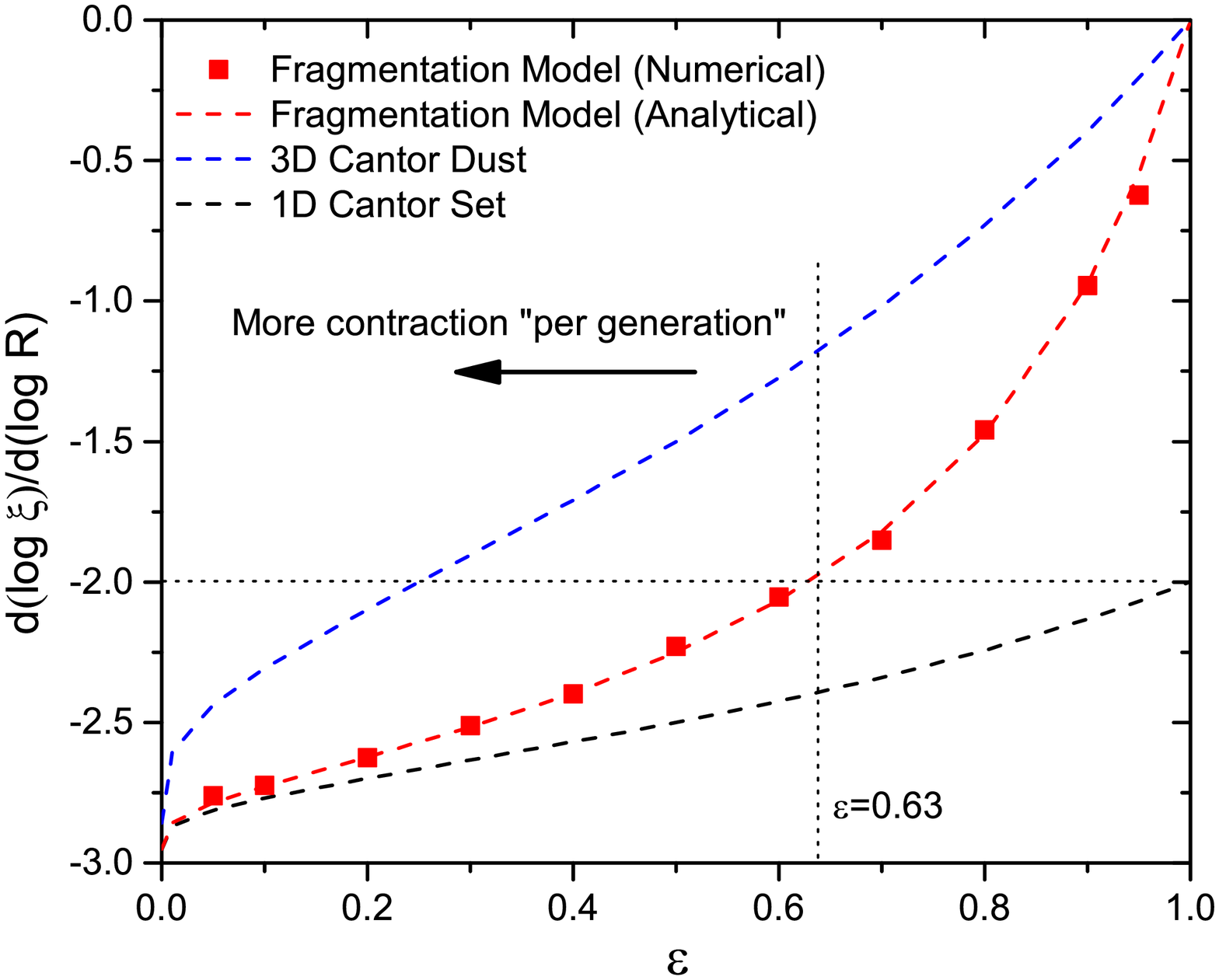}
\caption{Numerically calculated slopes of the 3D Cantor-like model (symbols) along with the analytical approximation (red) and the analytical result for the generalized 1D Cantor set (black) and 3D Cantor dust (blue).}
\label{fig:cantor_model_slope}
\end {center}
\end{figure}

We can analytically calculate the fractal dimension of our Cantor-like model if we take the separation between fragments to be the mean value of $R_{\rm parent}/2$. If we take the initial radius of the first sphere to be unity then, after $n$ iterations, the number of the objects is $N=2^{n}$ while their size is $R_n=\epsilon^N 2^{-N/3}$. If we choose a random fragment then the number of fragments within an $R_{m}$ radius is $N_m=2^{n-m}$, thus
\be
D\equiv \frac{\dderiv \ln{N_m}}{\dderiv \ln{R_m}}=\frac{\ln 2}{\frac{1}{3}\ln 2-\ln \epsilon}.
\label{eq:cantorl_slope}
\ee
Fig. \ref{fig:cantor_model_slope} shows that this result is actually exact. Since isothermal fragmentation is quite similar to this toy model, we expect the predicted stellar correlation function to have a slope between -1 and -3 (for reasonable $\epsilon$ values). This also implies that if some additional physics makes fragmentation harder (increasing the density threshold and thus decreasing $\epsilon$) then the correlation function becomes steeper.

This model has the free parameter $\epsilon$ which we can restrict by assuming that the fragmentation criteria is set by the Jeans-instability. It is known that the mass of fragments would be of the Jeans-mass $M_{\rm Jeans}\propto \rho^{-1/2}\propto \epsilon^{3/2}$. The number of fragments is $M/M_{\rm Jeans}=2$ which leads to $\epsilon=0.63$. For this value Eq. \ref{eq:cantorl_slope} gives $D=1$ leading to a $\xi\propto r^{-2}$ power law, in perfect agreement with our results from Fig. \ref{fig:allstar_corr_comp}.

\end{document}